\documentclass[8pt]{article}

\usepackage[table]{xcolor}

\usepackage{graphicx,pslatex,float,color,array,amssymb,amsmath,euscript}
\usepackage{pxfonts}
\usepackage{ifthen}
\usepackage{longtable}
\usepackage{eso-pic}
\usepackage{sidecap}
\usepackage{xspace}
\usepackage[paperwidth=21.0cm,paperheight=29.7cm,textwidth=16cm,textheight=23cm,top=3cm,left=2.5cm]{geometry}

\definecolor{navy}{rgb}{0,0,0.5}
\usepackage[pdftex,
           hyperindex=true,
           colorlinks=true,
           linkcolor=navy,
           anchorcolor=magenta,
           citecolor=navy,
           urlcolor=navy,
           unicode,
           implicit=true]{hyperref}

\usepackage{csquotes}

\usepackage{caption}

\usepackage{subcaption}
\usepackage{url}

\usepackage{soul,ulem,color,xspace,bm}
\renewcommand{\vec}[1]{\boldsymbol{#1}} 
\definecolor{darkgreen}{rgb}{0,0.5,0}
\newcommand{\be}{\begin{equation}}
\newcommand{\ee}{\end{equation}}

\title{Trapped fluid in contact interface}
\author{Andrei G. Shvarts, Vladislav A. Yastrebov\footnote{Corresponding author $\langle$\texttt{vladislav.yastrebov@mines-paristech.fr}$\rangle$}}

\date{\small{\it MINES ParisTech, PSL Research University, Centre des Mat\'eriaux}\\{\it CNRS UMR 7633, BP 87, F 91003 Evry, France}}

\begin{document}

\maketitle

\begin{flushleft}
 \Large{\bf Abstract.}\normalsize
\end{flushleft}

\noindent 	
	We study the mechanical contact between a deformable body with a wavy surface and a rigid flat taking into account pressurized fluid trapped in the interface. A finite element model is formulated for a general problem of trapped fluid for frictionless and frictional contact. Using this model we investigate the evolution of the real contact area, maximal frictional traction and global coefficient of friction under increasing external pressure. Elastic and elasto-plastic material models, compressible and incompressible fluid models and different geometrical characteristics of the 
wavy surface are used. 
	We show that in case of incompressible fluid, due to its pressurization, the real contact area and the global coefficient of friction decrease monotonically with the increasing external pressure. 
	Ultimately the contact opens and the fluid occupies the entire interface resulting in vanishing of static friction. In case of compressible fluids with pressure-dependent bulk modulus, we demonstrate a non-monotonous behaviour of the global coefficient of friction due to a competition between non-linear evolution of the contact area and of the fluid pressure. 	
However, for realistic compressibility of solids and fluids, contact-opening cannot be reached at realistic pressures. 
	An asymptotic analytical result for the trap-opening pressure is found and shown to be independent of the surface slope if it is small.
	On the other hand,
	in case of elastic-perfectly plastic materials, we 
again observe fluid permeation into the contact interface. 
	Finally, we study the distribution of frictional tractions during the depletion of the contact area under increasing pressure. 
	This process leads to emergence of singularity-like peaks in the tangential tractions (bounded by the Coulomb's limit) near the contact edges.
	We point out the similarity between the processes of trap opening and interfacial crack propagation, and estimate the complex stress intensity factor in the framework of linear elastic fracture mechanics.

\begin{flushleft}
\end{flushleft}

\section{Introduction}
\label{sec::intro}
The study of mechanical contact and friction is a subject of high importance in many fields, from biological and engineering applications to geological sciences. Since natural and industrial surfaces always possess roughness under certain magnification, the contact between solid bodies occurs on separate patches corresponding to asperities of contacting surfaces,~\cite{Archard_1953,Archard_1957,Bowden_2001,Greenwood_1966}. The evolution of the ratio of real contact area to apparent one under increasing external load determines essential contact properties such as friction, wear, adhesion, 
and is responsible for heat 
 transport 
 through contact interfaces. 
At the same time, the distribution of the free volume between contacting surfaces governs the fluid transport along the interface and is responsible for leakage/percolation phenomena, see for example~\cite{Dapp_2012,paggi2015evolution}.

\begin{figure}[t]
	\centering
	\begin{subfigure}{0.325\textwidth}
		\begin{center}
			\includegraphics[width=0.95\textwidth]{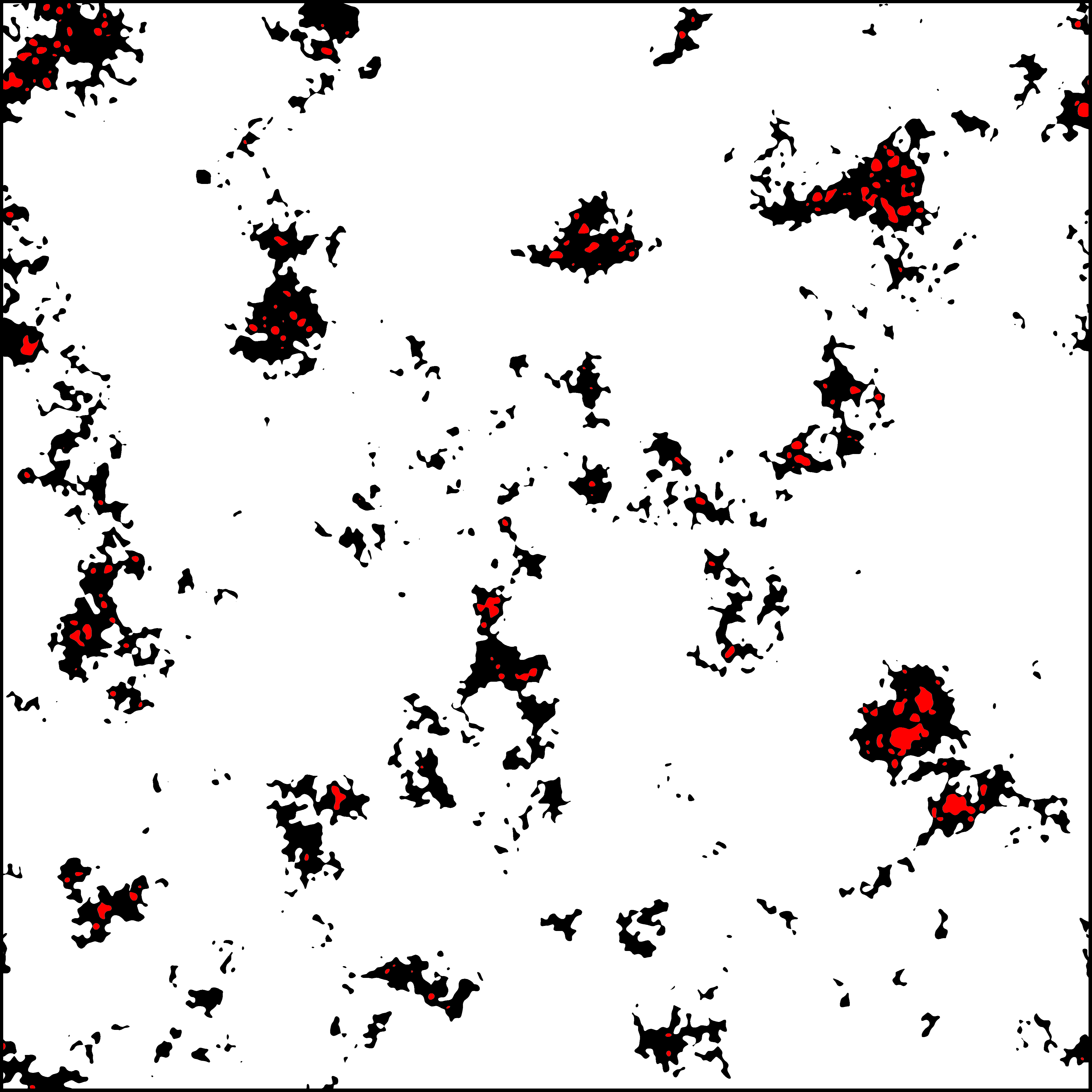} 
			\caption{ }
			\label{fig::trapped_area_1}
		\end{center}
	\end{subfigure}
	\begin{subfigure}{0.325\textwidth}
		\begin{center}		
			\includegraphics[width=0.95\textwidth]{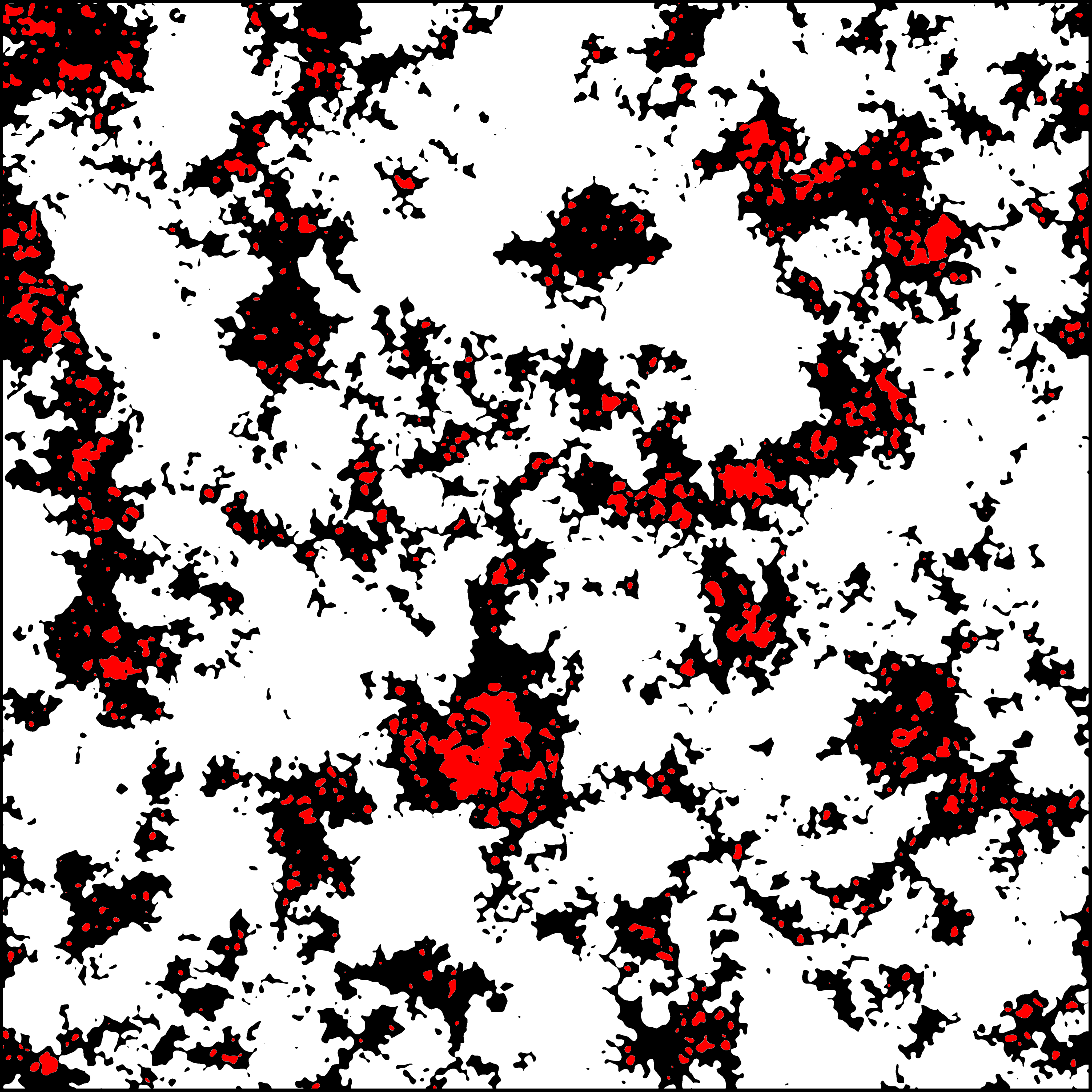} 
			\caption{ }
			\label{fig::trapped_area_2}
		\end{center}
	\end{subfigure}
	\begin{subfigure}{0.325\textwidth}
		\begin{center}		
			\includegraphics[width=0.95\textwidth]{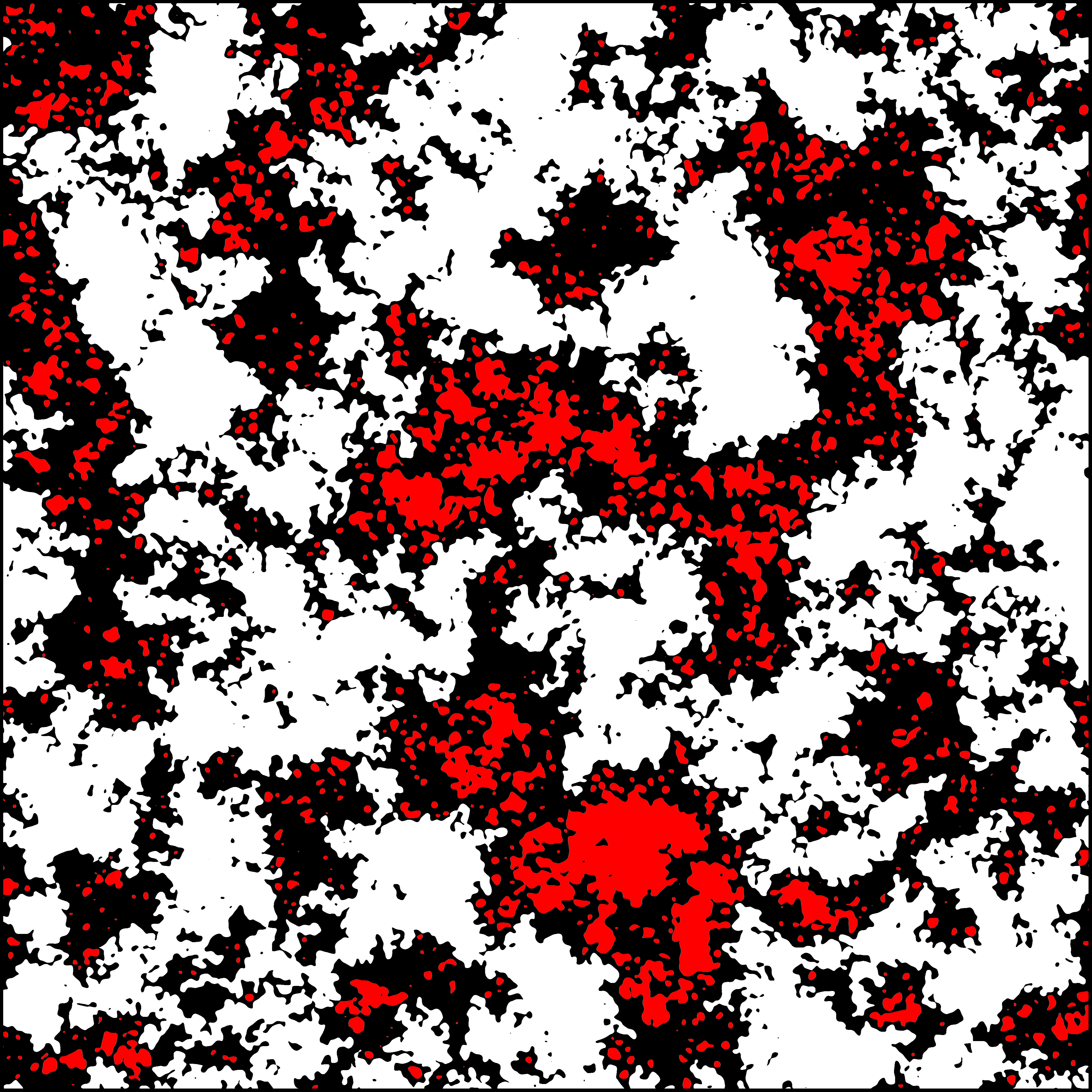} 
			\caption{ }
			\label{fig::trapped_area_3}
		\end{center}		
	\end{subfigure}
	\caption{Morphology of the contact interface between an elastic half-space with a rough surface and a rigid flat under increasing external load, numerical simulation results~\cite{Yastrebov_2015}: black is the real contact area, white is the ``free'' out-of-contact area and red is the \enquote{trapped} out-of-contact area, bounded inside non-simply connected contact patches.}
	\label{fig::trapped_area}	
\end{figure}

Lubrication, i.e. separation of 
contacting surfaces by a fluid lubricant, is an efficient mechanism for friction and wear reducing. However, if the applied external load, pushing the contacting bodies together, is high enough or if the sliding velocities are small, the hydrodynamic pressure developing in the fluid is not sufficient to separate the solids, and asperities of both surfaces can get in direct contact despite the presence of the lubricant, which inevitably increases friction. This scenario corresponds to the so-called mixed regime, at which the load-bearing capacity is split between the fluid and the contact areas. For even higher pressures and lower velocities, the whole load is carried by the mechanical contacts, this regime is termed as boundary lubrication, see~\cite{Hamrock_2004,Azushima_2016} for details. On the other hand, under increasing external load the lubricating fluid may be trapped in valleys (pools) delimited completely by the contact zone. Fig.~\ref{fig::trapped_area} shows an example of the morphology of the contact interface between two elastic half-spaces with rough surfaces under external load~\cite{pei2005finite,carbone2008asperity,putignano2012influence,yastrebov_2017}. Note that the fraction of the \enquote{trapped} out-of-contact area (highlighted by red color), surrounded by contact patches, is significant.

The entrapment of the fluid in the interface can have a strong effect on the contact properties, especially if the fluid is highly incompressible~\cite{persson2012elastic,Matsuda_2016}. First, the trapped fluid resists the compression, and thus opposes the growth of the real contact area. Second, the applied external load is shared between 
contacting asperities of the bodies and the pressurized fluid, so that the trapped fluid provides an additional load-carrying capacity (even in motionless contacts), reducing the normal pressure in the contact spots between the solid bodies. According to Coulomb's law of friction, the maximal tangential traction at the contact spots is proportional to the normal pressure, therefore the maximal macroscopic frictional force (of the whole contact interface) is proportional to the integral value of the normal pressure over the real contact area. Consequently, by taking into account the presence of the pressurized trapped fluid, a reduction of the global (apparent) coefficient of friction should take place.

The effect of lubricant entrapment on reduction of friction was first recognized in the study of cold metal forming processes \cite{Kudo_1965,Nellemann_1977}, where the authors performed experiments on the sheet metal drawing test and identified three states, corresponding to different levels of the external pressure~\cite{Azushima_1995}. Low values of external load are supported completely by the mechanical contact between asperities, and both global and local coefficients of friction are equal. At medium range of pressures, the global coefficient of friction decreases with increasing load due to closing of lubricant pools and generation of hydrostatic pressure in the fluid, which supports a part of the external load. At even higher load, fluid escapes from the pools and permeates into the contact zones, so that both the real contact area and the coefficient of friction decrease with increasing load. This effect is however biased by the fact that the real contact area does not evolve linearly under high pressures~\cite{Archard_1957}, but rather as a concave function of pressure~\cite{Persson_2001,yastrebov_2017}, thus also resulting in formal decrease of the friction coefficient in contact spots. Nevertheless, experimental results together with finite-element simulations of the problem of entrapment and permeation of the fluid into the contact interface during upsetting of a cylinder were presented and aforementioned states were also identified~\cite{Azushima_2000,Azushima_2011}. An extensive experimental study of lubricant entrapping and escape in cold rolling processes was presented in~\cite{Bech_1999}.

In biological sciences the effect of trapped lubricant in human joints was investigated in the view  of reduction of friction between rough cartilage surfaces~\cite{Soltz_2003, Chan_2011}. The concept of trapped fluid rises in the study of fatigue cracks in the rolling contact,  which considers the process of crack growth due to pressurized fluid lubricant, forced inside of the crack by the external load and trapped there~\cite{Bower_1988}. The trapped fluid problem is also relevant to the geophysical studies: a landslide or an earthquake can be caused by an elevation of the pressure of the fluid in the pores inside the rock, see for example~\cite{Viesca_2012,Garagash_2012}. 
The effect of the trapped fluid is also of interest for the study of basal sliding of glaciers,~\cite{Cuffey_2010}: the melt water, which is responsible for the lubrication, flows in a linked system of cavities in the interface between the glacier and the bedrock, and may be trapped there. Finally, the trapped fluid problem is also of importance for poromechanics~\cite{yu2002fractal,dormieux2002jmps,budiansky1976ijss,Coussy_2004}.

\cite{Kuznetsov_1985} extended the Westergaard's celebrated analytical solution for the problem of contact between a regular wavy surface and a rigid half-plane~\cite{Westergaard_1939} by taking into account the presence of a compressible fluid, trapped in the valleys between contacting asperities. Kuznetsov's solution demonstrates how the external pressure is divided between the fluid and the solid contact, which results in the decrease of the global coefficient of friction under increasing external load. 
However, due to 
the assumptions (i) that the wavy surface behaves as a flat one and that Flamant's solution holds for every surface point, and (ii) that the horizontal component of the fluid pressure is negligible, it cannot describe the escape of the lubricant and depletion of the real contact area. 
This question will be discussed later in detail.
Recently an analytical solution was proposed for the problem of sliding of a rigid periodical punch along a viscoelastic Winkler's foundation with the incompressible fluid present in the gap~\cite{Goryacheva_2012}.

Despite a significant attention to the problem of the trapped fluid in the contact interface, a few questions remain open, such as: the mechanism of the trap opening, the evolution of the real contact area and of the global coefficient of friction during this process, and also the distribution of the frictional shear tractions in the contact interface under external normal loading in the presence of the pressurized fluid in the interface. Note that these questions cannot be addressed in the framework of the boundary element method (BEM), since it assumes infinitesimal slopes of the surface roughness, which is, as we will show, a too restrictive assumption for the considered problem. We address these questions in the current study in the framework of the finite element method (FEM).

The paper is organized as follows. In Section~\ref{sec::problem} we present the statement of the problem of the mechanical contact coupled with pressurized compressible and incompressible fluids trapped in the contact interface. 
In Section~\ref{sec::analytic} we outline existing analytical solutions of this problem, and in Section~\ref{sec::methods} we discuss methods for its numerical solution. 
Section~\ref{sec::results} is devoted to results, including comparison of Kuznetsov's analytical solution with our numerical simulations, the evolution of the real contact area and the global coefficient of friction, as well as the simulation and analysis of the frictional behaviour of the system under normal and tangential external loading. In Section~\ref{sec::conclusions} we present the conclusions. 
Under the assumption of small slopes, in \ref{app:aux} we derive an analytical solution for vertical displacement of a wavy surface under the action of uniformly distributed pressure.
\ref{sec::appendix} provides details of the numerical formulation of the coupled trapped fluid/mechanical contact problem.

\section{Problem statement}
\label{sec::problem}

\begin{figure}[h]
	\centering
	\includegraphics[width=0.6\textwidth]{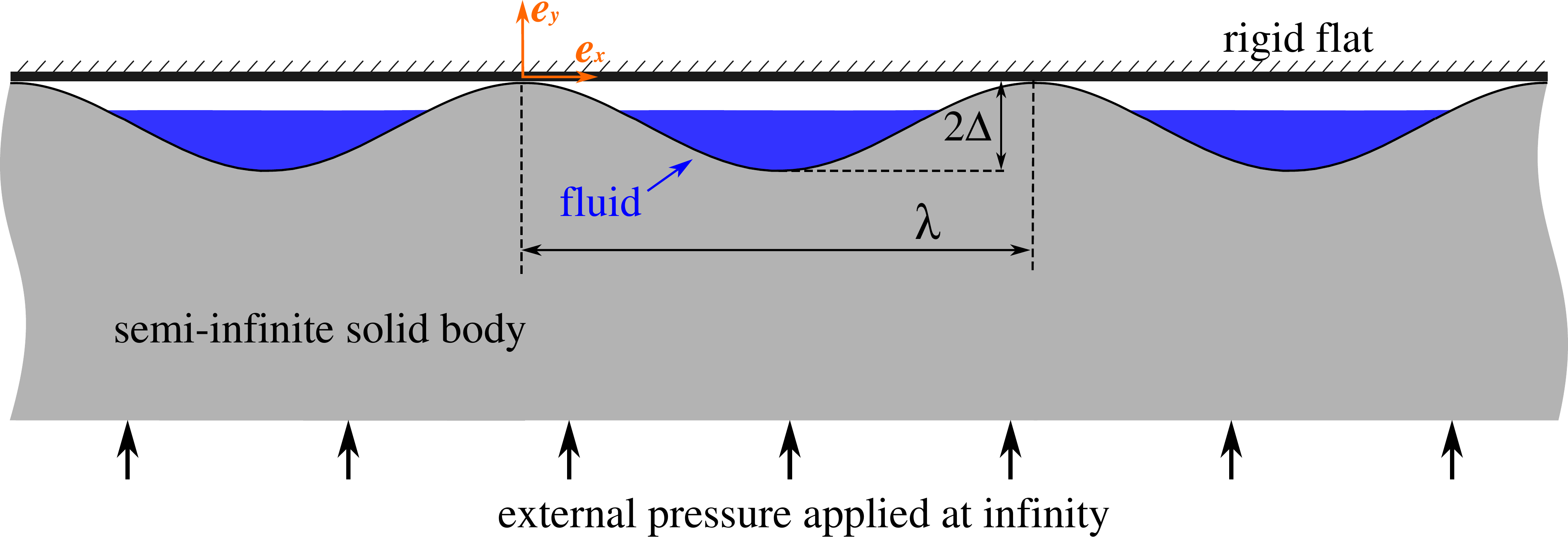} 
	\caption{A sketch of the problem under study.}
	\label{fig::problem}
\end{figure}

We consider a mechanical contact problem between a deformable half-plane with a periodic wavy surface and a rigid flat under the action of a far-field external pressure. This case was historically the starting point for the study of contact of rough surfaces~\cite{Westergaard_1939,Johnson_1985}. In addition, we take into account the influence of compressible or incompressible fluid trapped in the free volume between the two bodies, see Fig.~\ref{fig::problem}. We assume the plane strain problem and a linear elastic or elastic-perfectly \textit{J2-}plastic (the latter one is presented in Section~\ref{sec::elasto_plastic}) isotropic constitutive laws for the solid. Note that this problem is similar to the one already solved by~\cite{Kuznetsov_1985}, with the difference that we 
assume {\it small but finite} profile's slope, which, as will be shown in the paper, is of great importance for an accurate treatment of this problem.

The initial gap between the wavy profile and the rigid plane, as well as the volume of this gap for one wavelength of the profile, are given, respectively, by:
\begin{equation}
\label{eq::gap_volume}
g_0(\mathbf{X}) = \Delta \left(1 - \cos\frac{2 \pi \mathbf{X}}{\lambda}\right), \quad V_{g0} = l\:\int\limits_0^\lambda \Delta \left(1 - \cos\frac{2 \pi \mathbf{X}}{\lambda}\right) \; d\mathbf{X} = l\:\lambda\:\Delta,
\end{equation}
where $\Delta$ and $\lambda$ are the amplitude and wavelength of the wavy surface profile, respectively, $X$ is the horizontal coordinate in the initial (reference) configuration and $l$ is the length in the direction of the third coordinate $z$, which under the assumption of the plane strain state of deformation will be assumed hereinafter equal to one length unit.

\section{Analytical solutions\label{sec::analytic}}

\subsection{Westergaard's solution}

The problem of contact between an elastic half-space with a regular wavy surface $y=\Delta\cos(2\pi x/\lambda)$ and a rigid flat without fluid in the interface was solved by~\cite{Westergaard_1939}, see also~\cite{Johnson_1985}, for the case of infinitesimal ratio $\Delta / \lambda \ll 1$, i.e. infinitesimal slope of the roughness profile. According to this solution the pressure distribution inside contact patches ($-a + \lambda n \leq x \leq a + \lambda n, \: n \in \mathbb{Z}$) is given by:
\begin{equation}
p_W(x, a) =  \frac{2 \pi E}{1 - \nu^2}\frac{\Delta}{\lambda} \cos{\frac{\pi x}{\lambda}} \sqrt{\sin^2{\frac{\pi a}{\lambda}} - \sin^2{\frac{\pi x}{\lambda}}},
\label{eq::p_west}
\end{equation}
where $a$ is the half-length of contact patch within one wavelength of the profile $\lambda$, and elsewhere $p_W = 0$. $E$ and $\nu$ are Young's modulus and Poisson's ratio, respectively. The mean pressure over the whole contact interface is given by
\begin{equation}
\bar{p}_W(a) = \frac{1}{\lambda}\int\limits_0^{\lambda} p_W(x, a) \; dx = p^* \sin^2 \frac{\pi a}{\lambda},
\label{eq::p_west_mean}
\end{equation}
where $p^* = \pi E^* \Delta / \lambda$ is the pressure necessary to bring the entire interface in contact. In the static equilibrium $\bar{p}_W$ is equal to the value of the external pressure that we will denote by $p_{0}$. The complete contact is ensured, if $p_0 \geq p^*$.

By introducing the notations $A=2a$ and $A_0=\lambda$ for the real and apparent contact areas, respectively, the ratio of the real contact area to the apparent one, based on the Westergaard's solution, is given by:
\begin{equation}
\label{eq::area_west}
\frac{A}{A_0} = \frac{2a}{\lambda} = \frac{2}{\pi} \arcsin{\sqrt{\frac{p_0}{p^*}}}, \; 0\leq p_0\leq p^*.
\end{equation}

\subsection{Kuznetsov's solution}
\cite{Kuznetsov_1985} extended the Westergaard's solution~(\ref{eq::p_west}) by taking into account compressible fluid trapped in the valleys between contacting peaks of the wavy profile. Similarly, under the assumption of infinitesimal slope of the profile\footnote{
As was mentioned earlier, the infinitesimal-slope assumption implies here that (i) the wavy surface behaves as a flat one and that Flamant's solution holds for every surface point, and (ii) that the horizontal component of the fluid pressure is negligible.}, the stress state in the contact interface in the presence of the additional fluid pressure, applied beyond the contact patches, was considered as the superposition of the stress state corresponding to the same contact area, but without influence of the fluid (i.e. the Westergaard's solution~(\ref{eq::p_west})), and a uniform field of the fluid pressure $p_f$, applied everywhere and assumed not to distort the surface profile:
\begin{equation}
p_K(x, a) = \begin{cases}
p_f(a) + p_W(x, a), &\quad \text{if} -a + \lambda n \leq x \leq a + \lambda n, \: n \in \mathbb{Z}\\
p_f(a), &\quad \text{elsewhere}.
\end{cases}
\label{eq::kuz_press_eq}
\end{equation}
Integration of $p_K(x, a)$ over one period of the waviness gives the following relation between the external pressure $p_0$ and the contact area: $p_0(a) = p_f(a) + \bar{p}_W(a)$, where $\bar{p}_W(a)$ was defined in~\eqref{eq::p_west_mean}.

The fluid pressure $p_f$ can be related to the current contact half-width $a$ using a model of the compressible fluid with a bulk modulus $K$, which is defined as the ratio of infinitesimal pressure increase to the relative decrease of the volume:
\begin{equation}
K =  -V_f\frac{d p_f}{dV_f}.
\label{eq::comp_def}
\end{equation} 
In the linear compressibility model bulk modulus is a constant coefficient of proportionality between the relative change of volume of fluid and the fluid pressure~\cite{Kuznetsov_1985}:
\begin{equation}
p_f = K \left(1-\frac{V_f}{V_{f0}}\right),
\label{eq::comp_lin_model}
\end{equation} 
where $V_{f0}$ is the volume of the fluid in unpressurized state and a smaller volume $V_f$ corresponds to the fluid pressure $p_f$. However, the linear model of compressible fluid~(\ref{eq::comp_lin_model}) does not provide satisfactory results for most of the fluids used in real-life lubrication problems, since a significant dependence of the compressibility modulus $K$ of fluid on the pressure $p_f$ takes place~\cite{Kuznetsov_1985}. The simplest model, and yet quite precise for most of lubricating fluids, which takes into account this dependence is the compressibility linearly evolving with pressure~\cite{Nellemann_1977, Kuznetsov_1985}:
\begin{equation}
K = K_0 + K_1 p_f,
\label{eq::comp_modulus_nonlin}
\end{equation} 
where $K_0, K_1>0$ are model parameters. The linear dependence~(\ref{eq::comp_modulus_nonlin}), substituted into~(\ref{eq::comp_def}), upon integration results in the following non-linear relation between the fluid pressure and its volume:
\begin{equation}
p_f =  \frac{K_0}{K_1} \left\{\left(\frac{V_f}{V_{f0}}\right)^{-K_1}-1\right\}.
\label{eq::comp_nonlin}
\end{equation}	
Finally, it can be noted that the volume of the pressurized fluid $V_f$ is equal to the volume of the gap between the contacting surfaces $V_g$, which can be found from the displacement field of the Westergaard's solution~\cite{Kuznetsov_1985} and related to the current contact half-width $a$:
\begin{equation}
V_g(a) = V_{g0} \left[1 - \sin^2 \frac{\pi a}{\lambda} \left( 1 - \ln \left\{\sin^2 \frac{\pi a}{\lambda} \right\}\right)\right],
\label{eq::v_v0}
\end{equation}
where $V_{g0} = l \: \Delta$ is the initial gap, i.e., corresponding to $a=0$.

We generalize original results~\cite{Kuznetsov_1985} and allow a partial filling of the initial gap by the fluid, so that $V_{f0} = \theta \: V_{g0}, \; 0 < \theta \leq 1$. 
Therefore, if the current gap volume is bigger than the initial fluid volume, $V_g > V_{f0}$, i.e. $V_g/V_{g0} > \theta$, then the fluid is not yet pressurized, and Westergaards solution is valid: $p_{0}(a) = \frac{\pi E^* \Delta}{\lambda} \sin^2\frac{\pi a}{\lambda}$. If $V_g < V_{f0}$, or, equivalently, $V_g/V_{g0} <\theta$, the equation connecting the contact area and the external load has the following form in the case of linear compressible fluid:
\begin{equation}
\label{eq::p_kuz_lin}
p_{0}(a) = \frac{\pi E^* \Delta}{\lambda} \sin^2\frac{\pi a}{\lambda} + \frac{K}{\theta} \left[\theta - 1 + \sin^2 \frac{\pi a}{\lambda} \left(1 - \ln\left\{\sin^2 \frac{\pi a}{\lambda}\right\}\right)\right], \quad\mbox{ if } V_g/V_{g0} <\theta,
\end{equation}
and in the case of non-linearly compressible fluid:
\begin{equation}
\label{eq::p_kuz_nonlin}
p_{0}(a) = \frac{\pi E^* \Delta}{\lambda} \sin^2\frac{\pi a}{\lambda} + \frac{K_0}{K_1}\left[\theta^{K_1}\left(1 - \sin^2 \frac{\pi a}{\lambda} \left( 1 - \ln \left\{\sin^2 \frac{\pi a}{\lambda} \right\}\right)\right)^{-K_1}-1\right], \quad\mbox{ if } V_g/V_{g0} <\theta.
\end{equation}

It is important to note also that Kuznetsov's solution even in the case of an arbitrary large modulus of compressibility of the fluid shows the growth of the contact patches under the increasing load. Furthermore, in the limit of incompressible fluid $K \rightarrow \infty$ it gives a constant value of the real contact area, which can be found from the equation $V_g(a) = V_{f0}$. Consequently, Kuznetsov's solution, based on the assumption of infinitesimal slope of the profile, cannot predict depletion of the real contact area and escape of the fluid from the trap, which we demonstrate in following sections dropping out the assumption of infinitesimal slopes.

\section{
Numerical methods\label{sec::methods}}

\subsection{Mechanical contact}

In case of the unilateral contact between a deformable body and a rigid flat with an outer normal $\boldsymbol{\nu}$, the motion of the body is constrained, which can be formalized upon introduction of the normal gap function $g$ -- a signed distance from the points on the surface of the deformable body to the rigid plane:
\begin{itemize}
	\item[] $g > 0$, when the point is separated from the plane,
	\item[] $g < 0$, when the point penetrates the plane (which is forbidden),
	\item[] $g = 0$, when the point is on the plane.
\end{itemize}  
We will denote by $\Gamma$ the potential contact zone (the whole surface), by $\Gamma_c \subset \Gamma$ the active contact zone, where the normal surface traction $\sigma_n$ must be negative in non-adhesive contact, and by $\Gamma\setminus\Gamma_c$ the inactive zone, which is out of contact. The constraints governing the frictionless unilateral contact problem are known as the Hertz-Signorini-Moreau conditions~\cite{Wriggers_2006}:
\begin{equation}
\label{eq::contact_cond}
g \geq 0, \; \sigma_n \leq 0, \; \sigma_n \: g = 0 \quad \text{at} \; \Gamma \quad \Leftrightarrow \quad
\begin{cases}
g = 0, & \sigma_n < 0, \quad \text{at} \:\Gamma_c \\
g > 0, & \sigma_n = 0, \quad \text{at} \:\Gamma\setminus\Gamma_c.
\end{cases}
\end{equation}

Therefore the considered problem is the constrained minimization problem for the potential energy of the mechanical system $\Pi(\vec{u})$, where $\boldsymbol{u}$ is the displacement field.
This problem can be solved using the Lagrange multipliers method~\cite{Kikuchi_1988,Wriggers_2006}, with the Lagrangian functional defined as:
\begin{equation}
\label{eq::Lagrangian_contact}
\mathcal{L}(\mathbf{u}, \lambda_c) = \Pi(\mathbf{u}) + \int\limits_{\Gamma_c} \lambda_c \, g(\mathbf{u}) \: d\Gamma_c,
\end{equation}
where $\lambda_c \leq 0$ is the Lagrange multiplier function, the values of which are equivalent to the normal traction in the contact zone. 

\begin{figure}[h]
	\centering
	\begin{subfigure}{0.49\textwidth}
		\centering
		\includegraphics[width=0.99\textwidth]{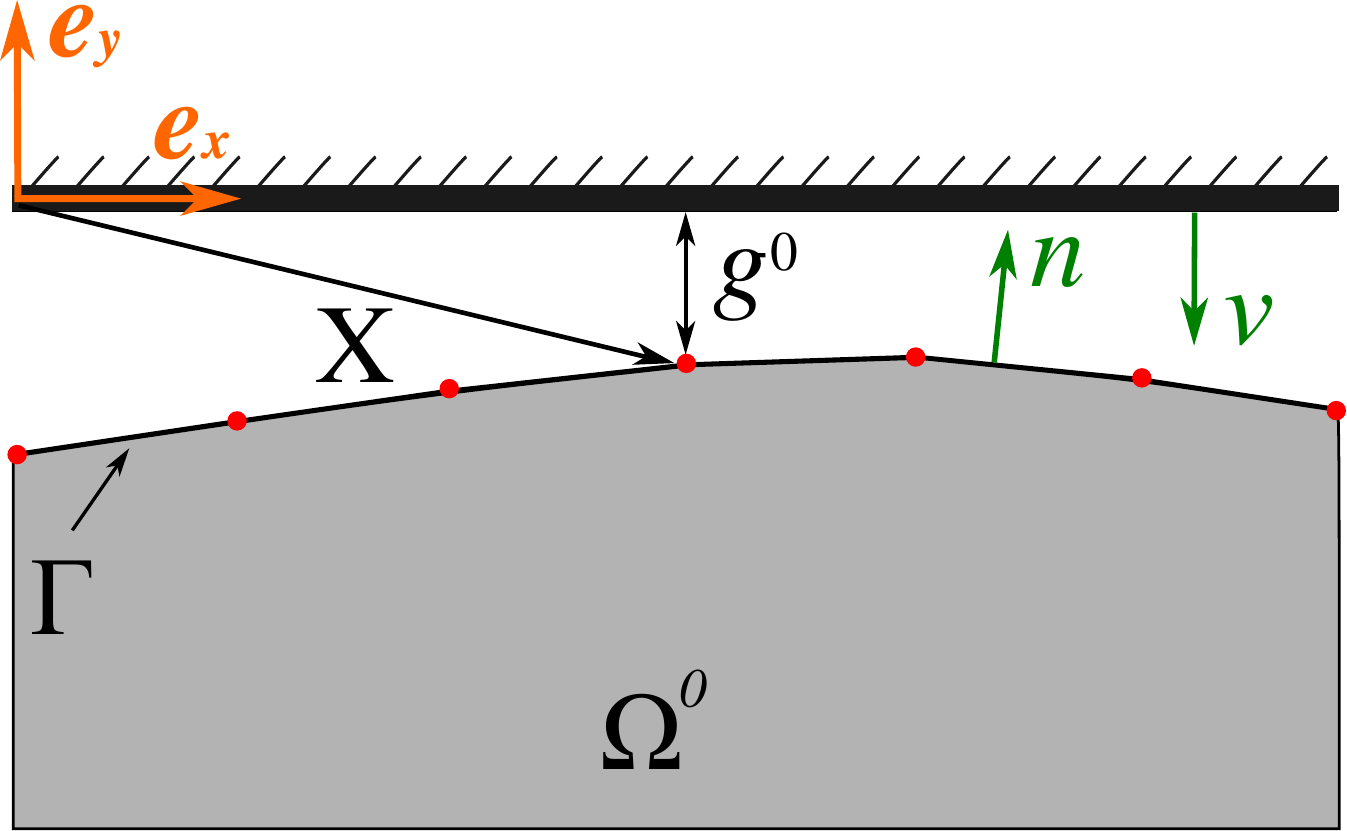} 
		\caption{ }
		\label{fig::inactive_contact}
	\end{subfigure}\hfill
	\begin{subfigure}{0.49\textwidth}
		\centering
		\includegraphics[width=0.99\textwidth]{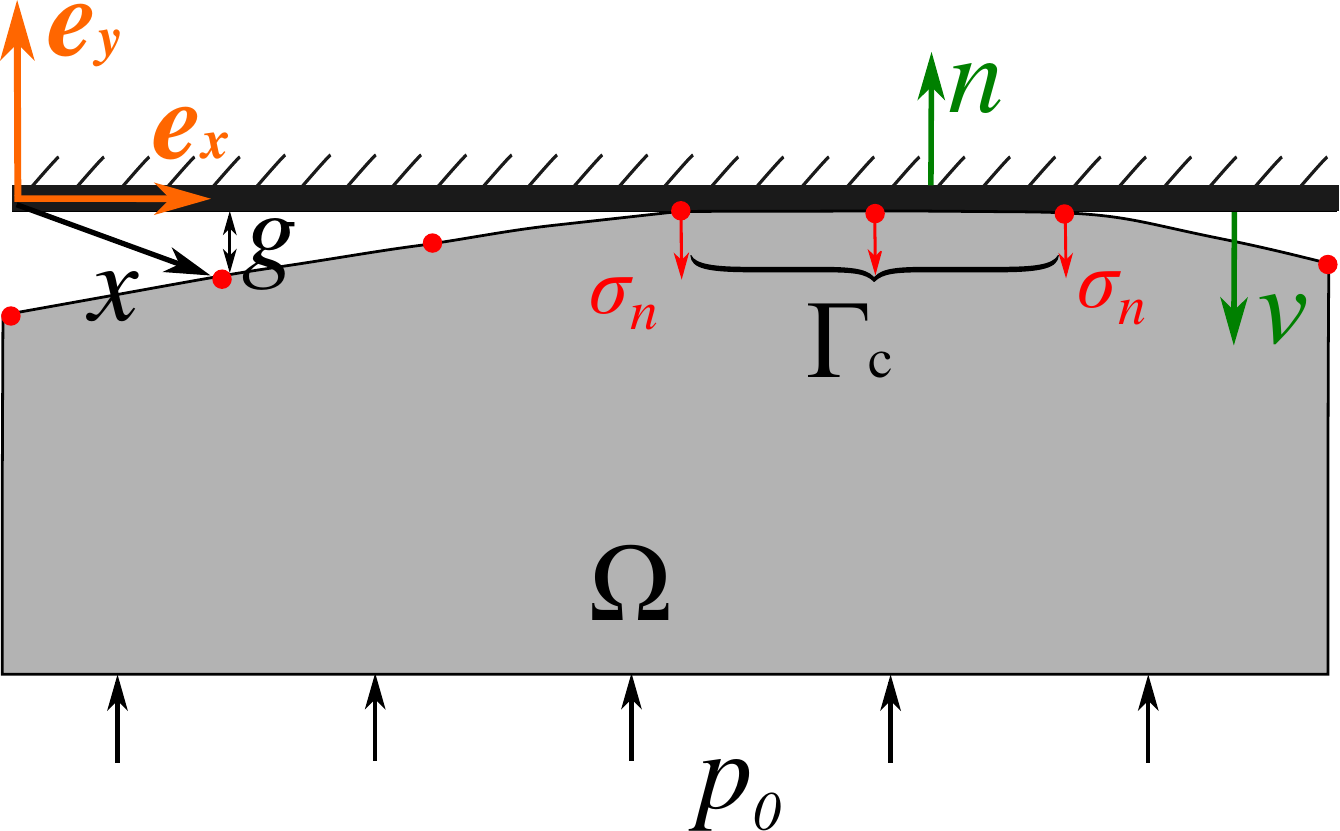} 
		\caption{ }
		\label{fig::active_contact}
	\end{subfigure}
	\caption{(a) Reference configuration: $\boldsymbol{X} \in \Omega^0$. (b) Actual configuration: $\boldsymbol{x} \in \Omega$, $p_0$ is the external pressure.}
\end{figure}

In case of frictional contact, along with Hertz-Signorini-Moreau  conditions~(\ref{eq::contact_cond}), additional frictional constraints must be included in the problem, such as Coulomb's law of friction, which defines the following possible active contact states:
\begin{itemize}
	\item Stick: $\vec{\dot g}_t = 0, \; |\vec\sigma_t| < \mu\: |\sigma_n|$;	
	\item Slip: $\vec\sigma_t = \mu \: |\sigma_n|\; \vec{\dot g}_t/|\vec{\dot g}_t|$;
\end{itemize}
where $\vec{\dot g}_t$ is the sliding velocity in the tangential plane between the corresponding points of the two surfaces, $\vec{\sigma}_t$ is the tangential contact traction and $\mu$ is the coefficient of friction (CoF).

In order to include frictional constraints in the formulated above constrained minimization problem, special methods must be used, such as the penalty method (combined with the return mapping algorithm) or augmented Lagrangian method, for details see~\cite{Wriggers_2006,Yastrebov_2013}.

\subsection{Trapped fluid constraints}

\subsubsection{Geometrical constraint for incompressible fluid}

\begin{figure}[h]
	\centering
	\begin{subfigure}{0.49\textwidth}
		\centering
		\includegraphics[width=0.99\textwidth]{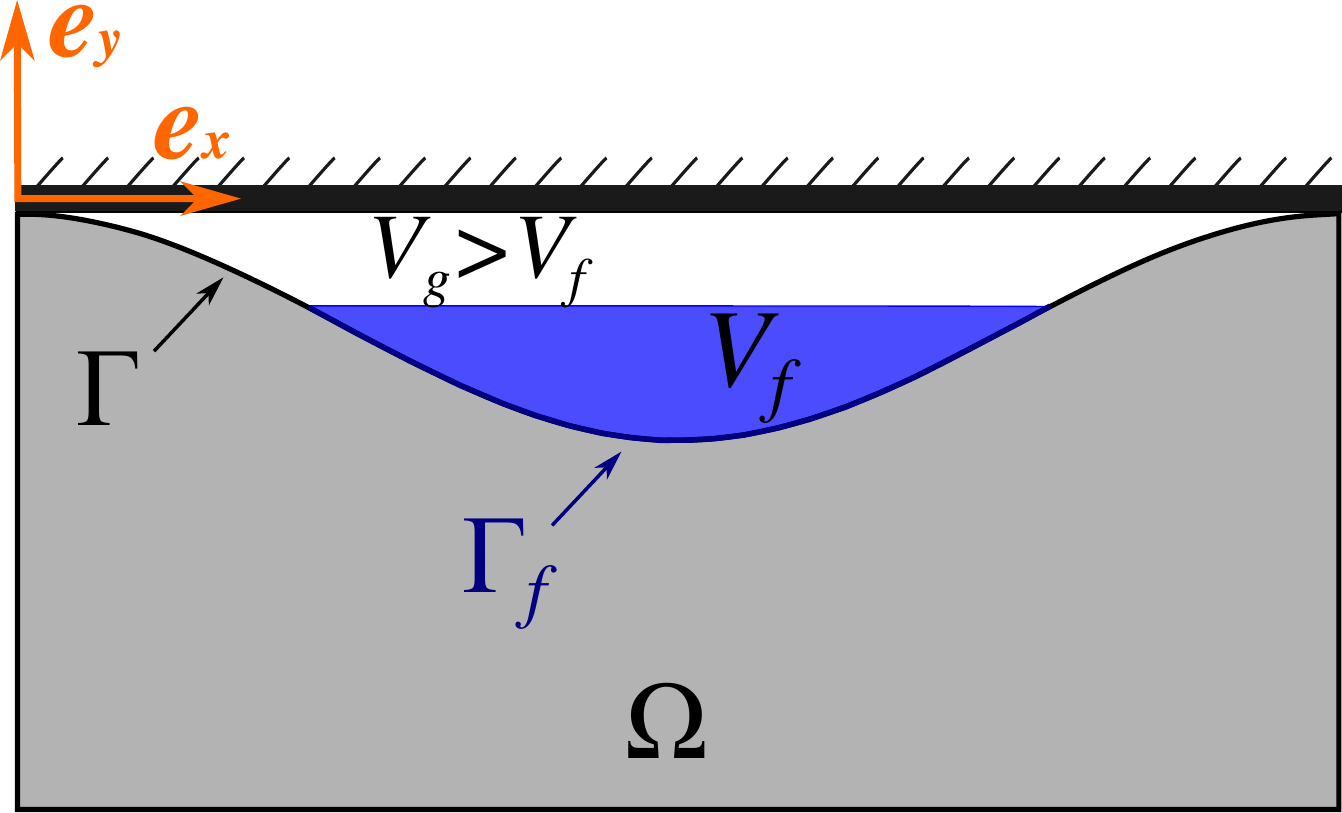} 
		\caption{ }
		\label{fig::inactive}
	\end{subfigure}\hfill
	\begin{subfigure}{0.49\textwidth}
		\centering
		\includegraphics[width=0.99\textwidth]{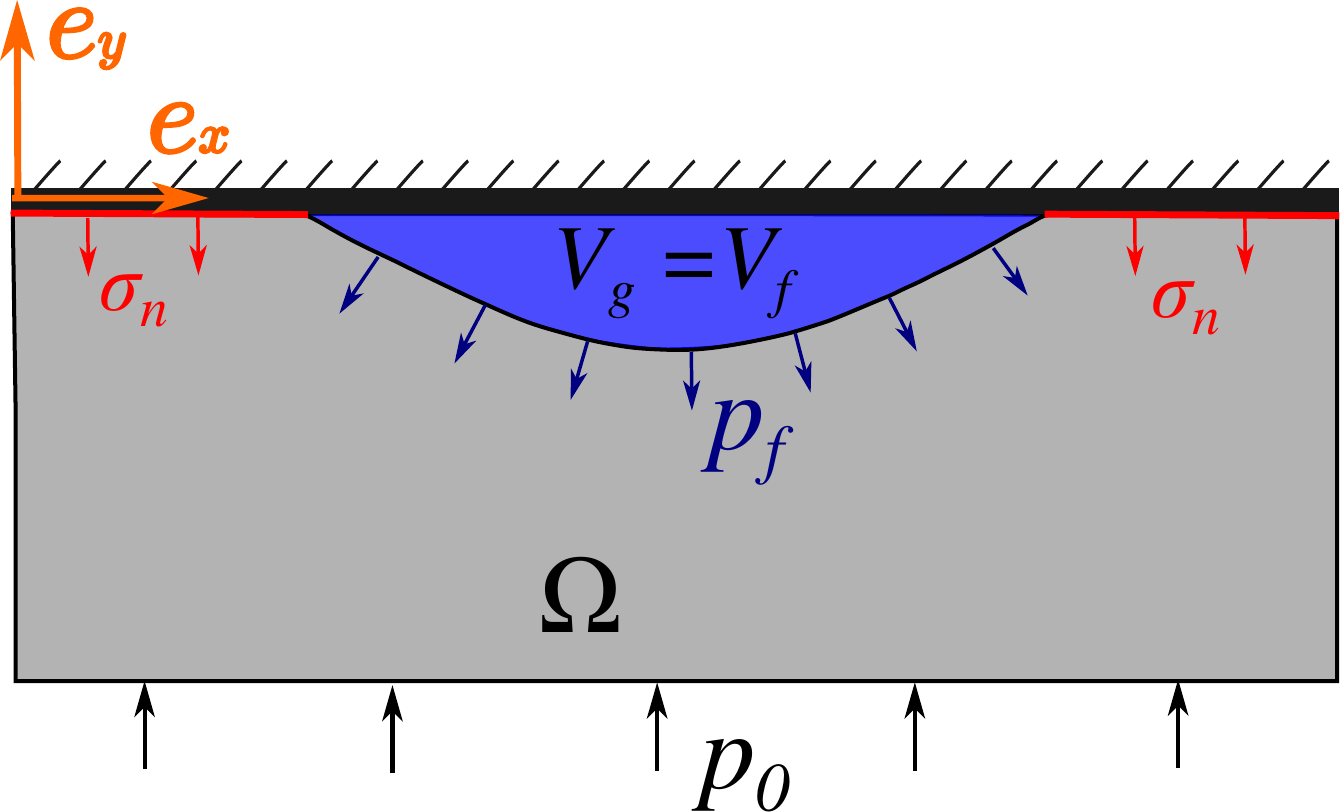} 
		\caption{ }
		\label{fig::active}
	\end{subfigure}
	\caption{(a) Trapped fluid is in inactive state. (b) Trapped fluid is in active state, $p_f$ is the fluid pressure.}
\end{figure}

The area of the gap between the contacting surfaces $V_g$ in the presence of trapped incompressible fluid of volume $V_f$ must satisfy the following geometrical constraint: 
\begin{equation}
\label{eq::gap_volume_integral}
V_g \geq V_f = \text{const}, \quad V_g(\boldsymbol{X}+\mathbf{u}) = \int\limits_{\widetilde{\Gamma}_f}g(\boldsymbol{X}+\mathbf{u}) \: d\widetilde{\Gamma}_f,
\end{equation}
where $\Gamma_f = \Gamma \setminus \Gamma_c$ and $\widetilde{\Gamma}_f$  is the projection of $\Gamma_f$ on the rigid plane. The trapped fluid may fill completely or partially the gap between the contacting surfaces, therefore it can be present in two different states: ``inactive'', when $V_f < V_g$ and the fluid if not pressurized ($p_f$ = 0), and ``active'', when $V_f = V_g$, and pressure in the fluid $p_f>0$, see Fig.~\ref{fig::inactive},~\ref{fig::active}. We may formulate this two states in a way similar to Hertz-Signorini-Moreau conditions:
\begin{equation}
\label{eq::fluid_cond}
V_g \geq V_f, \quad p_f \geq 0, \quad p_f \: (V_g - V_f) = 0 \Leftrightarrow 
\begin{cases}
V_g = V_f, & p_f > 0, \quad \text{(active state)}\\
V_g > V_f, & p_f = 0, \quad \text{(inactive state)}.
\end{cases}
\end{equation}

\subsubsection{Simulation of  incompressible fluid using a Lagrange multiplier}

In the case of the inactive state of the trapped fluid we have only the mechanical contact problem between the elastic body and the rigid plane, while if the fluid is in the active state, we must consider additionally the gap volume constraint~(\ref{eq::gap_volume_integral}).  The Lagrange multiplier method may be used again in order to fulfil this constraint, and the combined functional for the coupled problem can be defined as: 
\begin{equation}
\label{eq::Lagrangian_coupled}
\mathcal{L}(\mathbf{u}, \lambda_c, \lambda_f) = \Pi(\mathbf{u}) + \int\limits_{\Gamma_c} \lambda_c \, g(\mathbf{u}) \: d\Gamma_c - \lambda_f (V_g(\mathbf{u}) - V_f),
\end{equation}
where $\lambda_f \geq 0$ is the Lagrange multiplier for the trapped fluid problem, which is equivalent to the fluid pressure $p_f$. The solution of the coupled problem is a stationary point of the Lagrangian~(\ref{eq::Lagrangian_coupled}), which requires the calculation of its variation:
\begin{align}
\label{eq::Lagrangian_variation}
\delta \mathcal{L}(\mathbf{u}, \lambda_c, \lambda_f) =& \frac{\partial \Pi(\mathbf{u})}{\partial\mathbf{u}} \delta \mathbf{u} + \int\limits_{\Gamma_c} \left[\delta \lambda_c \; g(\mathbf{u}) + \lambda_c \;  \frac{\partial g(\mathbf{u})}{\partial\mathbf{u}} \delta \mathbf{u}\right] \; d\Gamma_c  \nonumber \\
-&\left[\delta \lambda_f \; (V_g(\mathbf{u}) - V_f) + \lambda_f \; \frac{\partial V_g(\mathbf{u})}{\partial\mathbf{u}} \delta \mathbf{u} \right]= 0.
\end{align}

\subsubsection{Simulation of the compressible fluid with the penalty method}

The geometrical constrain~(\ref{eq::gap_volume_integral}) for the trapped fluid can also be treated with the penalty method. In accordance with the linear penalty method, instead of the term $\lambda_f (V_g(\mathbf{u}) - V_f)$, the following term should be added in~(\ref{eq::Lagrangian_coupled}) to take into account the trapped fluid constraint:
\begin{equation}
W_f(\mathbf{u}) = \frac{\epsilon}{2} \: \left(V_{f0}-V_g(\mathbf{u}) \right)^2,
\label{eq::penalty_func}
\end{equation}
if the fluid is in active state $V_g < V_{f0}$, and zero otherwise. In the above formula $\epsilon$ is the penalty parameter, and $V_{f0}$ is the initial volume of the fluid.

Let us assume that the fluid is in active state. Calculating the variation of~(\ref{eq::penalty_func}), we obtain the contribution of the trapped fluid to the balance of virtual works: 
\begin{equation}
\delta W_f(\mathbf{u}) = -\epsilon \: (V_{f0}-V_g(\mathbf{u})) \frac{\partial V_g(\mathbf{u})}{\partial \mathbf{u}} \delta \mathbf{u},
\label{eq::penalty_func_var}
\end{equation}
where the value of the term $\epsilon \: (V_{f0}-V_g(\mathbf{u}))$ equals to the fluid pressure $p_f$. Under the penalty formulation the gap volume constraint~(\ref{eq::gap_volume_integral}) is never satisfied exactly, i.e. the current volume of the active fluid $V_g(\mathbf{u})$ is always smaller, than the initial fluid volume $V_{f0}$. Therefore, the penalty method corresponds to the model of the compressible fluid, and a comparison between~(\ref{eq::comp_lin_model}) and~(\ref{eq::penalty_func_var}) shows that the linear penalty method represents the  compressibility model with the constant bulk modulus $K$, if $\epsilon = K / V_{f0}$.

In order to simulate the behaviour of the compressible fluid with pressure-dependent bulk modulus~\eqref{eq::comp_modulus_nonlin}-\eqref{eq::comp_nonlin}, the \textit{non-linear penalty} method for the trapped fluid constraint~(\ref{eq::gap_volume_integral}) may be used. The contribution of the fluid to the balance of virtual works in this case takes the form:
\begin{equation}
\delta W_f = -\frac{K_0}{K_1}\left\{\left(\frac{V_g(\mathbf{u})}{V_{f0}}\right)^{-K_1}-1\right\}\frac{\partial V_g(\mathbf{u})}{\partial \mathbf{u}} \delta \mathbf{u}.
\label{eq::penalty_nonlin_var}
\end{equation}

\section{Results and discussion\label{sec::results}}

\begin{figure}[h]
	\centering
	\includegraphics[width=0.4\textwidth]{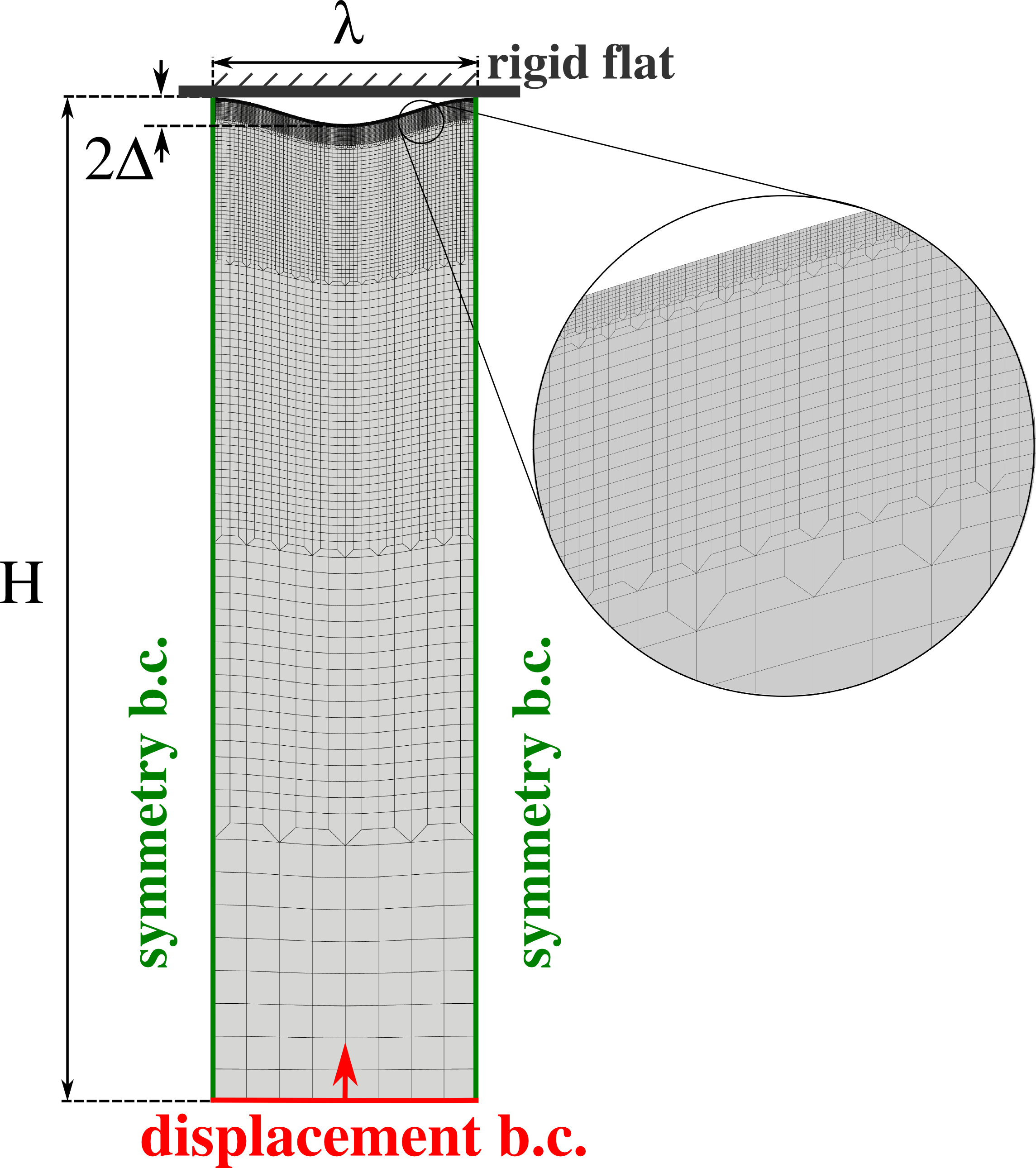} 
	\caption{FEM mesh}
	\label{fig::mesh}
\end{figure}

We solved the coupled problem using the finite element method with implemented monolithic coupling scheme in finite element suite \textit{Z-set}~\cite{Besson_1997,Zset}.
Contrary to 
Kuznetsov's analytical results or BEM analyses, we did not assume infinitesimal slopes, i.e. the value $\Delta / \lambda$ is arbitrary. We used a finite element mesh with $1024$ nodes in the contact interface per wavelength ($19364$ nodes in total in the structural mesh), see Fig.~\ref{fig::mesh}. Hereinafter, if not mentioned differently, we considered the roughness profile with $\Delta / \lambda = 0.01$. In the following, we will also discuss how this ratio affects the results. The horizontal dimension of the finite element mesh equals to 
 the wavelength $\lambda$ and the ratio of the profile amplitude $\Delta$ to the vertical mesh dimension $H$ is $\Delta / H = 0.005$. On the vertical boundaries of the mesh we apply symmetry boundary conditions ($u_x = 0$), and the bottom edge of the deformable solid is displaced vertically towards the rigid flat within 200 load steps. A corotational updated Lagrangian framework was used in our simulations, which is needed to capture properly that the fluid pressure applied to the updated configuration is collinear to element normals.
In simulation we measure the vertical reaction, the extension of the contact area, the pressure in the contact zone and the fluid pressure. 

Hereinafter, if not mentioned differently, we performed frictionless simulations, and estimated the value of the global coefficient of friction using the following approach. 
We consider the global coefficient of friction as the coefficient of proportionality between the maximal tangential force per wavelength $F_t$ and the normal one $F_n = p_0\lambda$, i.e.  $|F_t| \le \mu_{\mbox{\tiny glob}} |F_n|$. The local Coulomb's coefficient of friction determines the following inequality: $|\sigma_t| \leq \mu_{\mbox{\tiny loc}} |\sigma_n|$, where $\sigma_t$ and $\sigma_n$ are the tangential and normal components of the traction vector in the contact interface, respectively.

We neglect shear forces in the fluid and therefore the ratio between the global and local coefficients of friction can be calculated as:
\begin{equation}
\frac{\mu_{\mbox{\tiny glob}}}{\mu_{\mbox{\tiny loc}}} = \int\limits_{\Gamma_c}\left.\vphantom{A^A_A}|\sigma_n| \; d\Gamma_c \right/|F_n| = 
	\int\limits_{\Gamma_c}\left.\vphantom{A^A_A}|\sigma_n| \; d\Gamma_c \right/p_0 \lambda.
\label{eq::est}
\end{equation}
Finally, using notations for the real $A$ and the apparent $A_0$ contact areas, which were introduced above, Eq.~\eqref{eq::est} can be rewritten as: 
\begin{equation}
\frac{\mu_{\mbox{\tiny glob}}}{\mu_{\mbox{\tiny loc}}} = 1 - \frac{p_f}{p_0}\left(1-\frac{A}{A_0}\right).
\label{eq::est3}
\end{equation}

\subsection{Incompressible fluid\label{sec::results_incomp}} 

In this section we study the model of an incompressible fluid trapped in the contact interface. Note that real-life lubricating fluids have significantly lower initial bulk moduli than metals. Nevertheless, this idealized model enables us to focus on the mechanism of the trap opening by the pressurized fluid, while compressible fluids will be considered in the following sections.

\begin{figure}[h]
	\centering
	\includegraphics[width=1.0\textwidth]{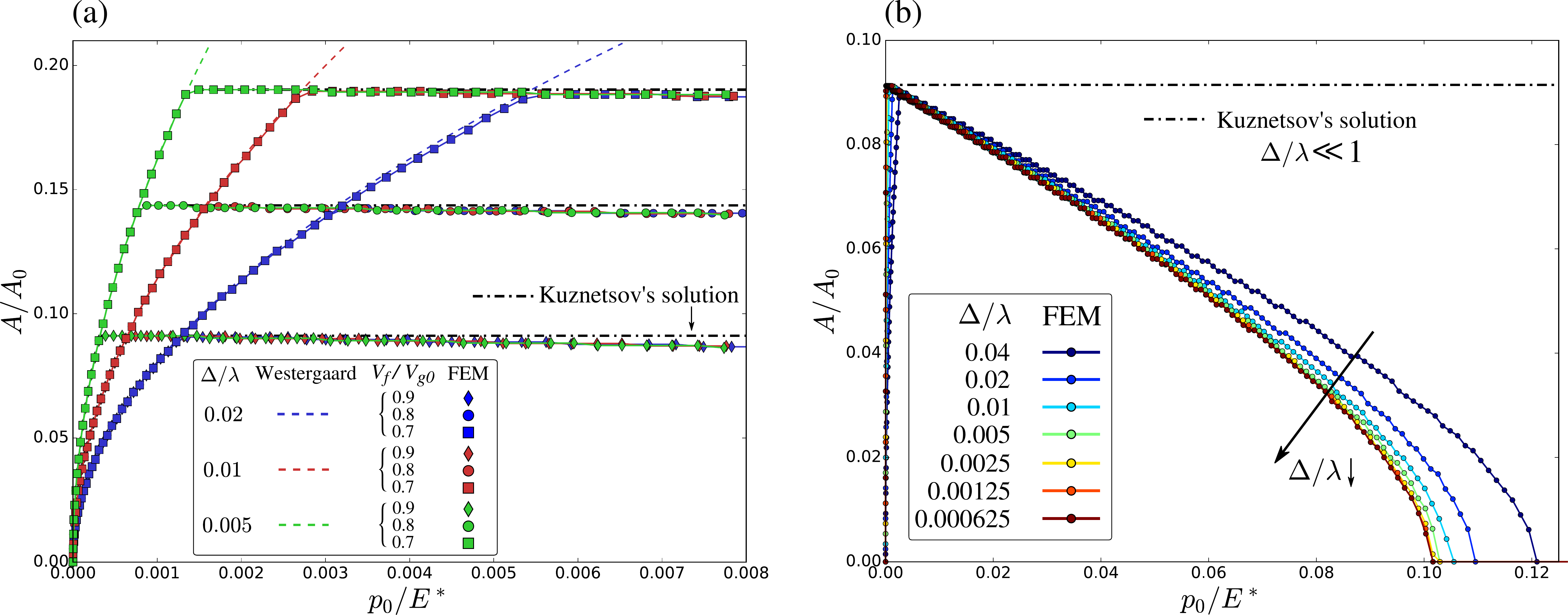} 
	\caption{
		(a) The evolution of the real contact area in the vicinity of the ``activation'' point of the incompressible fluid with respect to the external pressure normalized by $E^*$ for three profiles with different slopes $\Delta/\lambda$ and three cases with different ratios of the fluid volume to the initial gap volume $V_f/V_{g0}$. (b) The evolution of the real contact area until the complete opening of the trap for the case $V_f/V_{g0} = 0.9$ shown for different slopes $\Delta/\lambda$.
		\label{fig::incomp_2}
	}
\end{figure}

\begin{figure}[h]
	\centering
	\includegraphics[width=1.0\textwidth]{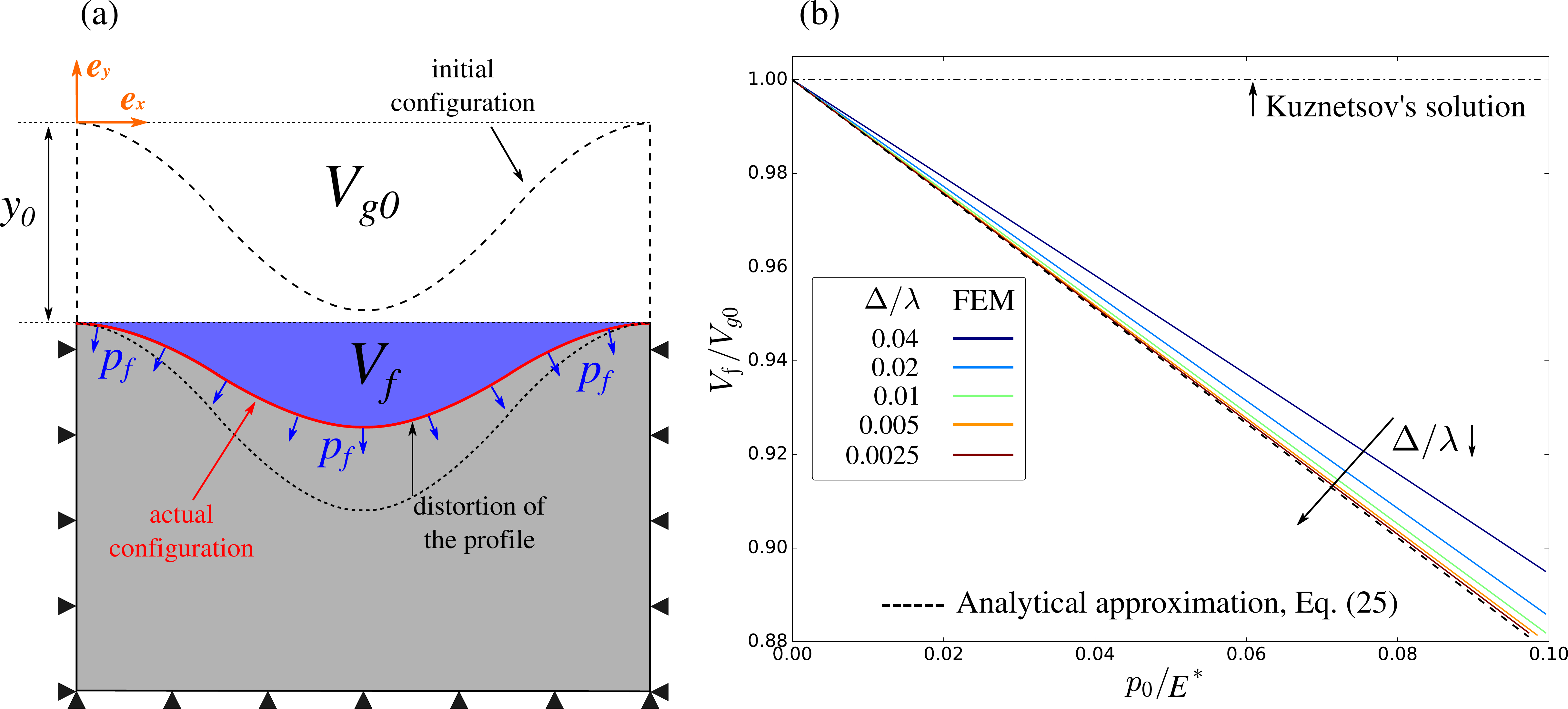} 
	\caption{(a) Sketch of the auxiliary problem: 
		deformation of the wavy surface under uniform normal load. (b) Evolution of the ratio $V_f/V_{g0}$ ($V_f$ is the volume between the deformed surface and a horizontal plane $y=y_0$, where $y_0$ is the current position of the crest, and $V_{g0}$ is the initial volume of the gap) with the increasing external pressure $p_0$ for several profiles with different slope $\Delta/\lambda$.}
	\label{fig::aux}	
\end{figure}

\begin{figure}[p]
	\centering
	\includegraphics[width=0.5\textwidth]{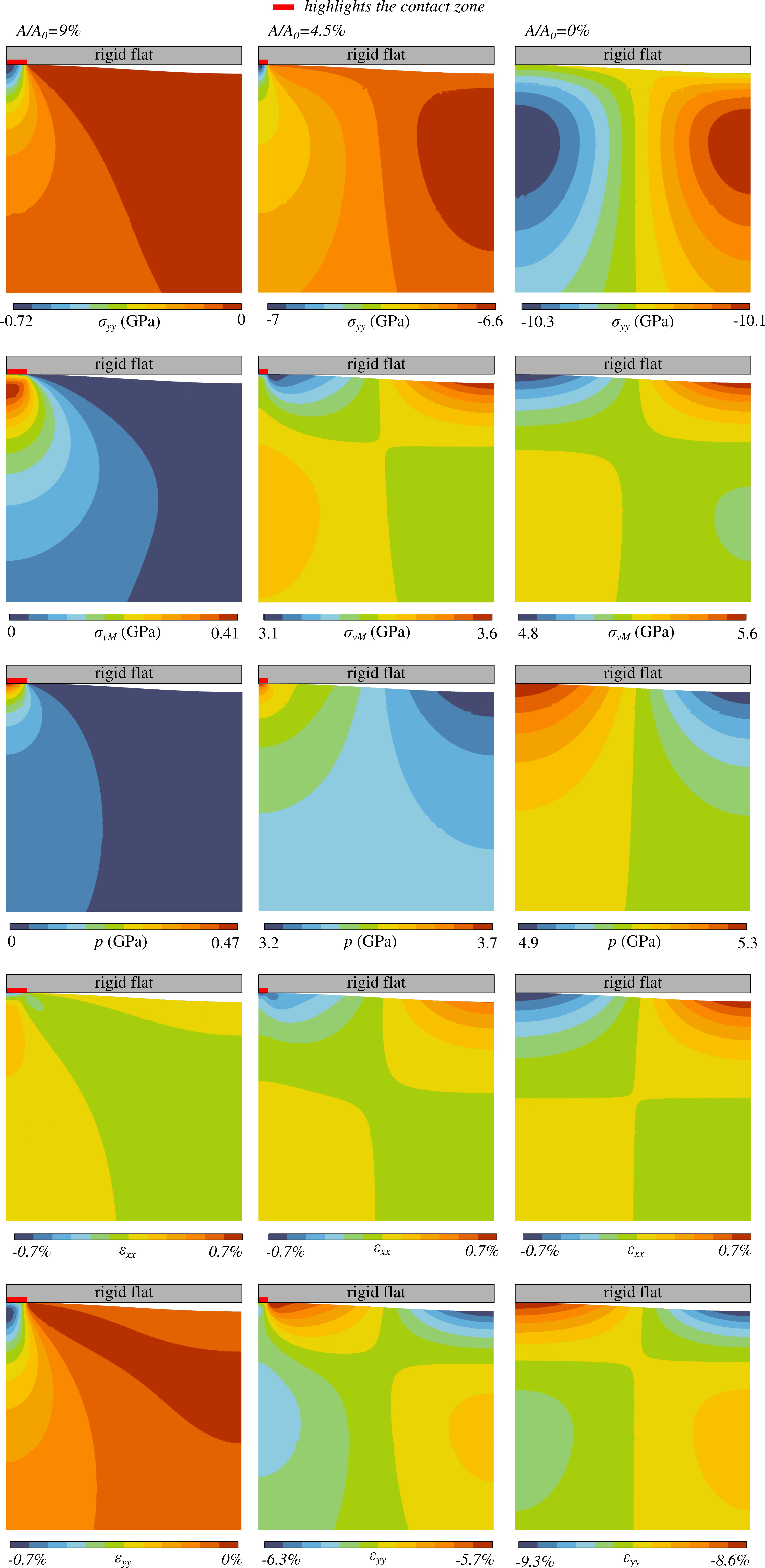} 
	\caption{Stress and strain components in the bulk of the deformable solid during the process of trap opening due to the increasing pressure in the fluid. 
		Top to bottom: vertical stress component $\sigma_{yy}$, von Mises stress $\sigma_{vM}$, hydrostatic stress $p$, horizontal strain component $\varepsilon_{xx}$ and the vertical one
		$\varepsilon_{yy}$. Three loading steps are considered, corresponding to, left to right: maximal contact area (activation of the fluid), half of the contact opened, contact area
		is zero (trap is opened). The considered elastic material is typical aluminium ($E = 70 \text{ GPa}, \nu = 0.33$), the fluid is assumed incompressible.
		\label{fig::el_map}
	}
\end{figure}

\begin{figure}[p]
\centering
\includegraphics[width=0.99\textwidth]{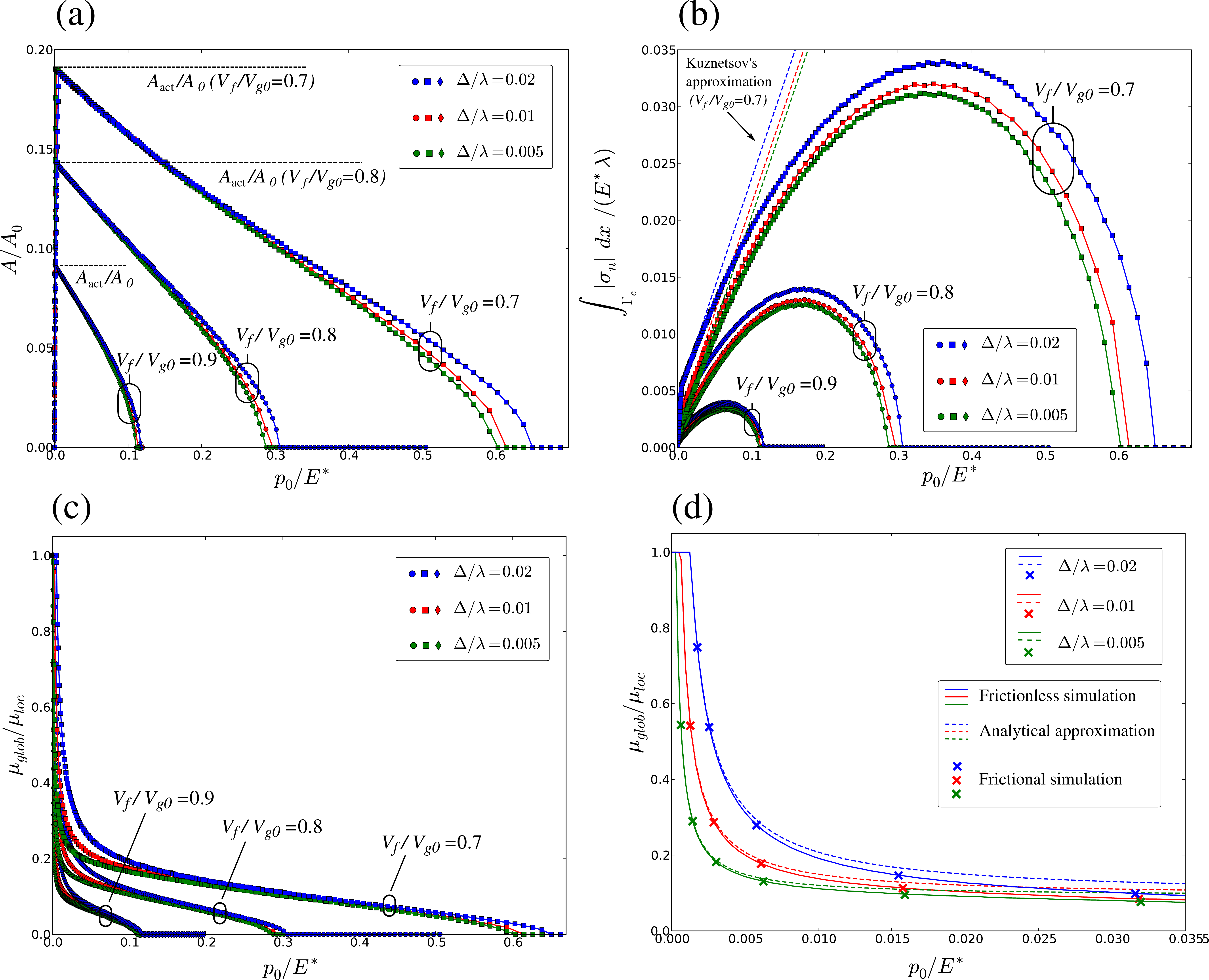} 
\caption{(a) Real contact area evolution during opening of the contact, caused by pressurized incompressible trapped fluid (with respect to the external pressure normalized by $E^*$) 
(b) Contact normal force evolution during opening of the contact (with respect to the external pressure, normalized by $E^*$). (c) Evolution of the ratio between global and local 
 coefficients of friction, and (d) a zoom of this evolution for $V_f/V_{g0} = 0.9$, where, in addition, the results of \emph{frictional} simulations are plotted (crosses), as well as analytical approximations given by~\eqref{eq::cof_kuz} (dashed curves).}
\label{fig::incomp}
\end{figure}

We study the evolution of the real contact area in the presence of incompressible fluid in the interface under the increasing external pressure using the Lagrange multiplier method. We investigate how the magnitude of the slope of the profile ($\Delta / \lambda$) and the ratio between the trapped fluid volume and the initial gap volume $V_f / V_{g0}$ affect the solution of the coupled problem. The distribution of some stress and strain components in the bulk of the deformable solid during the process of trap opening is shown in Fig.~\ref{fig::el_map}.

The evolution of the contact area close to the moment of the activation of the fluid is presented in Fig.~\ref{fig::incomp_2}(a). The regime in which the fluid is not yet pressurized ($V_g>V_f$) coincides with Westergaard's equation~\eqref{eq::p_west}. According to this analytical solution the ratio of the current volume of the gap to the initial one $V_g/V_{g0}$ is a monotonically decreasing function of contact area and does not depend on the slope of the profile $\Delta/\lambda$, see~\eqref{eq::v_v0}. Therefore, the contact area $A_\text{act}$, reached when the fluid gets pressurized ($V_g = V_f$) does not depend on the slope of the profile and is increasing with decreasing $V_f / V_{g0}$. For a given $\Delta/\lambda$ the pressure necessary to activate the fluid $p_{\text{act}}$ is also increasing with decreasing $V_f / V_{g0}$. At the same time, for a given $V_f / V_{g0}$, the value of $p_{\text{act}}$ is proportional to the slope $\Delta/\lambda$.

One can note in Fig.~\ref{fig::incomp_2}(a) that once the fluid is pressurized, the contact area is slowly decreasing, contrary to the Kuznetsov's solution, which predicts the contact area to remain constant. In Fig.~\ref{fig::incomp_2}(b) we show the evolution of the contact area in a much wider range of loads, than in Fig.~\ref{fig::incomp_2}(a), and observe a monotonic decrease of the contact area, ultimately it reaches zero value, which corresponds to the opening of the trap. Surprisingly, results of simulations with different (decreasing) profile slope $\Delta/\lambda$ do not tend to the Kuznetsov's solution (derived under assumption of infinitesimal $\Delta/\lambda$ and assuming that the wave profile is similar to a flat one), but converge to a different limit! At the same time we observe that the external pressure necessary to open the trap $p_\text{open}$ also converges to a certain limit with $\Delta/\lambda\to0$.

In order to explain this intriguing result, first we note that since the solution of linearly elastic problem with and without contact is unique, the displacement field at the moment of opening of the trap with $p_0 = p_\text{open}$ must be equal (up to a rigid body motion) to the one corresponding to distributed hydrostatic pressure $p_f = p_\text{open}$ over the whole interface. Let us consider an auxiliary problem of the uniform hydrostatic fluid pressure on the wavy profile, see Fig.~\ref{fig::aux}(a). The Kuznetsov's solution is based on an assumption that a uniform distribution of the hydrostatic pressure does not distort the wavy surface~\cite{Kuznetsov_1985}. In our numerical simulations we showed that for small, but finite $\Delta/\lambda$ this assumption does not hold, the wavy surface distorts: the crest's displacement is bigger than the displacement of the trough, which is quite an evident result. 

Due to the non-zero slope of the contact interface, the fluid pressure acts not only in the vertical direction but also in the horizontal one, thus leading to the additional in-plane compression of the material near the crest and, on the opposite, to the additional in-plane tensile contribution near the trough, see Fig.~\ref{fig::aux}(a). Thus, there exists a linearly elastic solution for a uniformly distributed pressure $p_f$, which results in such surface deformation, that the integral of the gap equals to the fluid volume $V_f$, i.e.:
\begin{equation}
\exists p_f\quad\text{such that}\int\limits_{\Gamma} (y_0 - (X^y+ u^y))d\Gamma = V_f,
\end{equation}
where $y_0$ is the position of the crest after the applying the uniform pressure $p_f$. 
We derived an analytical formula for computation of $V_f$, based on the assumption of small, but finite $\Delta/\lambda$:
\begin{equation}
V_f/V_{g0} = 1 - \frac{2(1-2\nu)(1+\nu)p_f}{E},
\label{eq:open_vol}
\end{equation}
see~\ref{app:aux} for details. The relative change of volume induced by a uniformly applied pressure $p_f$ does not depend on the value $\Delta/\lambda$, but only on elastic properties of the solid.
In Fig.~\ref{fig::aux}(b) the comparison of this formula with the numerical results for several profiles with different $\Delta/\lambda$ is shown. Numerical results are tending towards the analytical solution for decreasing $\Delta/\lambda$. Therefore, we have shown that for any given $V_f/V_{g0}$ there exists uniform pressure $p_f$, which results in a such distortion of the surface, that the volume between the surface and a rigid flat equals to $V_f$. Moreover, in the limit of infinitesimal slopes, this critical pressure does not depend on the slope.
	
The obtained result explains why the curves of evolution of the real contact area with the increasing pressure for surfaces with different slopes tend to a certain limit with decreasing $\Delta/\lambda$ (which remains, however, finite) see Fig.~\ref{fig::incomp_2}(b), while $p_\text{open}$ is different for different $V_f/V_{g0}$, see Fig.~\ref{fig::incomp}(a). Equation~\eqref{eq:open_vol} can be readily used to compute the pressure needed to open the trap; it is valid for incompressible or compressible fluids.

We present in Fig.~\ref{fig::incomp}(b) the evolution of the integral contact pressure, i.e. the nominator in~\eqref{eq::est}, for different values of $\Delta/\lambda$ and $V_f / V_{g0}$. The results show, that just after the fluid becomes pressurized, the integral of contact pressure has an almost linear growth, which follows the linear dependence of the contact reaction on the external pressure $p_0$, provided by the Kuznetsov's solution in the limit $K\rightarrow \infty$:
\begin{equation}
\label{eq::react_kuz}
\frac{1}{E^* \lambda}\int_{\Gamma_c}|\sigma_n| \; d\Gamma_c = \frac{p_0}{E^*} \frac{A_\text{act}}{A_0} + \pi \left(1 - \frac{A_\text{act}}{A_0}\right) \frac{\Delta}{\lambda} \sin^2{\frac{\pi}{2}\frac{A_\text{act}}{A_0}},
\end{equation}
where, contrary to numerical results, it was assumed that $A_\text{act}$ remains constant under the increasing external pressure $p_0$.

However, due to the fact that we consider finite slope of the profile in the numerical solution, the linear part in the dependence of contact reaction on external pressure is followed by a non-linear concave part, reaching maximum value and then decreasing to zero. Consequently, the global coefficient of friction also vanishes. The results on the estimation of the ratio between global and local coefficients of friction are presented in Fig.~\ref{fig::incomp}(c). 
Before the fluid gets pressurized, the global CoF equals to the local one. After that, the global CoF is monotonically decreasing with the increasing external pressure $p_0$. This decrease is related to repartition of the external load between the contact and the fluid; the latter is assumed not to resist shear in the quasi-static limit. Note that for high values of $p_0$, i.e. close to opening of the trap, the evolution of the global CoF is independent from the slope ($\Delta/\lambda$) and depends only on the ratio $V_f/V_{g0}$. On the other hand, for low values of $p_0$ slightly higher than the activation pressure (see Fig.~\ref{fig::incomp}(d)) the analytical approximation under the assumption of infinite $K$ shows the global CoF decreasing as $1/p_0$:
\begin{equation}
\label{eq::cof_kuz}
\frac{\mu_{\mbox{\tiny glob}}}{\mu_{\mbox{\tiny loc}}} = \frac{A_\text{act}}{A_0} + \pi \left(1 - \frac{A_\text{act}}{A_0}\right) \frac{\Delta}{\lambda} \frac{E^*}{p_0} \sin^2{\frac{\pi}{2}\frac{A_\text{act}}{A_0}}.
\end{equation}
Note, that the term containing $1/p_0$ is proportional to the ratio $\Delta/\lambda$.

In addition to estimations of the global coefficient of friction~\eqref{eq::est}-\eqref{eq::est3}, based on the frictionless simulation of the coupled problem under normal loading, we performed the direct computation of $\mu_{\text{glob}} = |F_t| / |F_n|$ in the frictional simulation of the coupled problem during sliding under normal and tangential loads. Note that in the latter simulation for both normal and frictional contact constraints we use the augmented Lagrangian method and the classic Lagrange multiplier method for the fluid constraint. The comparison of the results is presented in Fig.~\ref{fig::incomp}(d) for the case of $V_f/V_{g0} = 0.9$ and different ratios of $\Delta/\lambda$: the analytical asymptotic solution~\eqref{eq::cof_kuz} is presented with dashed curves, estimations based on frictionless simulation are shown as solid curves, the results calculated with taking into account friction in the interface are presented as crosses for a few particular values of external pressure $p_0$. This comparison shows that the frictionless result, based on the assumption of separate consideration of tangential and normal contributions in the interface~\cite{Johnson_1985}, provides a trustworthy estimation of the global coefficient of friction. 

Note that these considerations can be applied to multi-cracked materials such as rocks with fluid in contact interfaces.
The irreversible deformation in rocks is related to the frictional sliding at crack interfaces, 
which starts after the mean shear traction $\langle\sigma_t\rangle$ in the interface reaches the frictional limit determined by the coefficient of friction and the contact pressure $\mu_{\mbox{\tiny glob}}\langle\sigma_n\rangle$. Being homogenized over all randomly oriented crack orientations, these considerations give rise to Drucker-Prager-type constitutive behaviour with the initial yield surface given by $f = \sigma_{vm} + \mu_{\mbox{\tiny glob}} p - R_0$, where $\sigma_{vm}$ is the von Mises stress, $p = -\mathrm{trace}(\vec \sigma)/3$ is the hydrostatic pressure and $R_0$ is the initial yield stress for pure shear. 
Because of the presence of an incompressible fluid in the interface, the frictional limit does not increase linearly (or equivalently the global coefficient of friction does not remain constant), but reaches its maximum and decreases down to zero as shown in Fig.~\ref{fig::incomp}(b). This behaviour is very similar to advanced pressure-dependent plasticity models with a so-called cap, which corresponds to the decay of the von Mises yield stress with increasing pressure~\cite{resende1985formulation}. 
But contrary to the pore-collapse mechanism~\cite{suarez1990indentation,perrin1993rudnicki,issen2000conditions}, here this decay results from the decrease of the global friction with the hydrostatic pressure in presence of the fluid, this result also holds for non-linearly compressible fluids.

\subsection{Compressible fluid with constant bulk modulus\label{sec::results_comp_lin}}

\begin{figure}[h]
\centering
\includegraphics[width=0.99\textwidth]{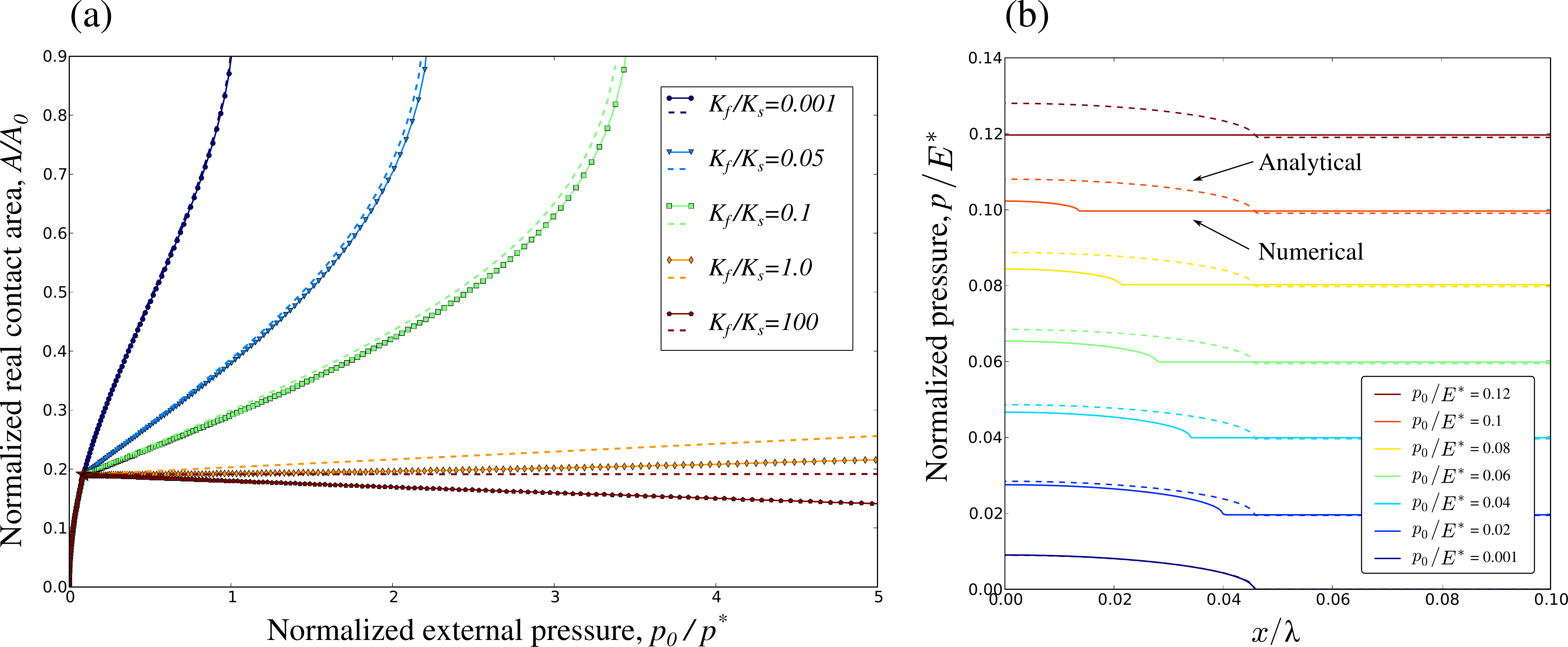} 
\caption{(a) Evolution of the ratio of the real contact area to the apparent one under increasing external pressure $p_0$: comparison of numerical and analytical results for different values of the fluid modulus of compressibility, normalized by the bulk modulus of the solid $K_f/K_s$; $\Delta/\lambda = 0.01, \: V_{f0}/V_{g0} = 0.7$. 
	(b) Distribution of the normal pressure near the contact patch under the increasing external load $p_0$. Solid lines are the results of the numerical simulation and dashed lines correspond to the analytical solution under the same external pressure, $\Delta/\lambda = 0.01, \: V_{f0}/V_{g0} = 0.9$, $K_f / K_s = 6 \cdot 10^4$.}
\label{fig::comp_lin}
\end{figure}

Here our analysis is extended to the case of compressible fluids.In Fig.~\ref{fig::comp_lin}(a) we present the comparison of the numerical simulation of a linearly compressible trapped fluid under the linear penalty formulation with the analytical solution \eqref{eq::p_kuz_lin}. We plot the evolution of the ratio of the real contact area to the apparent one under increasing external pressure for the case when the fluid occupies 70\% of the initial gap, i.e. $V_{f0}/V_{g0} = 0.7$. Different curves correspond to different values of the modulus of compressibility of the fluid $K_f$, normalized by the bulk modulus of the solid body $K_s = E / 3(1 - 2\nu)$, and for each numerical result a 
corresponding analytical curve is presented for comparison. 

Before pressurization of the fluid, the presence of the latter does not affect the solution and all curves follow the Westergaard's solution~\eqref{eq::p_west}. For the pressurized fluid, the results show a good agreement between numerical and analytical solutions for values $K_f / K_s \ll 1 $, and for $K_f \approx 0$ the solution coincides completely with the Westergaard's formula. 
However, with the increase of the $K_f$, in the region corresponding to the active fluid, the difference between numerical and analytical solutions becomes more pronounced. 
For the ratio $K_f/K_s$ close to unity, the numerical results shows an almost constant value of the real contact area under the increasing load. Note, that the same result will hold for an incompressible fluid trapped in the interface between two incompressible solids. 

For even greater $K_f/K_s$, the numerical results show a decrease of the real contact area, which means that the pressurized fluid starts to open the contact.
Due to inherent assumptions of infinitesimal slopes, these effects cannot be predicted by the analytical solution.

In Fig.~\ref{fig::comp_lin}(a) the results were presented for $V_f/V_{g0} = 0.7$, note that the smaller this ratio is, the bigger are the value of pressure necessary to bring the fluid in active state and the corresponding value of the contact area. However, after the fluid becomes pressurized, for sufficiently high values of external pressure, the evolution of the contact area is influenced only by the compressibility modulus of the fluid and the mean slope of the profile. The bigger is the compressibility modulus or the slope, the smaller is the contact area for the same external pressure. 

To emphasize the difference between the analytical and numerical solutions for a nearly incompressible fluid, we plot the pressure distribution near a contact patch under the increasing load for both solutions, see Fig.~\ref{fig::comp_lin}(b).
The representation of the stress state in the contact patches as a superposition of the stress state for the same contact area without the influence of the fluid and a uniform fluid pressure~(\ref{eq::kuz_press_eq}) still holds for the numerical solution, but unlike the analytic solution, in our results a significant reduction of the contact area for nearly incompressible fluid is observed. 

Note that in our numerical solution for sufficiently high external pressure the real contact area vanishes, which means that the fluid separates the contacting surfaces everywhere, and the external pressure is entirely supported by the fluid under the pressure equal to the external one $p_f=p_0$.

\subsection{Compressible fluid with pressure-dependent bulk modulus\label{sec:results_comp_nonlin}}

\begin{figure}[p]
\centering
\includegraphics[width=0.99\textwidth]{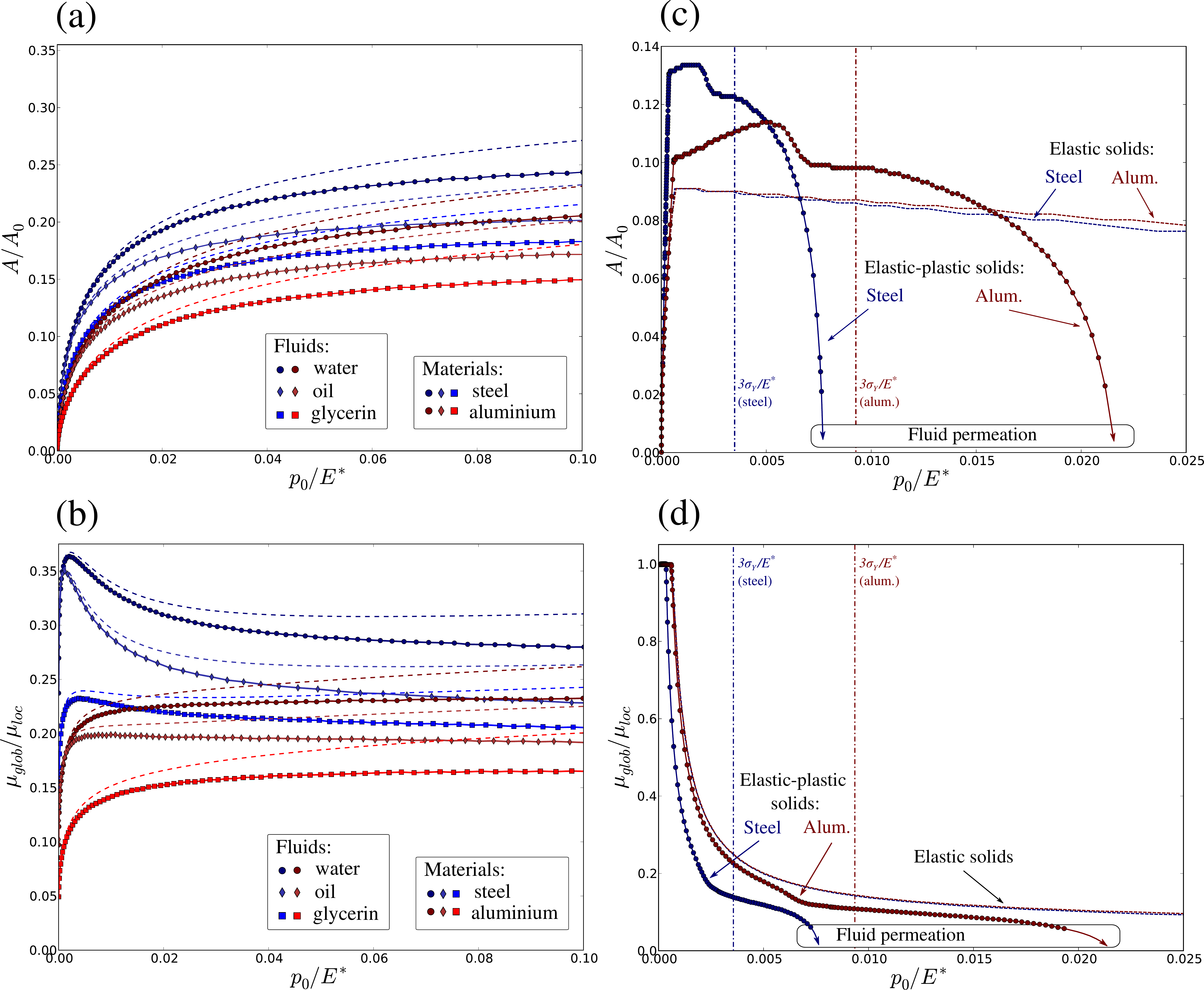} 
\caption{
Evolution of (a) the ratio of real contact area to the apparent one, (b) the ratio between global and local coefficients of friction under increasing external pressure 
for two elastic solids representing steel and aluminium, and non-linearly compressible fluids representing water, glycerine and oil. 
The dashed curves correspond to the analytical solution given by~\eqref{eq::p_kuz_nonlin}. 
Evolution of (c) the ratio of real contact area to the apparent one $A/A_0$, and of (d) the global to local coefficients of friction under increasing external pressure in the case of elastic-perfectly plastic solid and incompressible fluid. Note that in the initial configuration the fluid does not occupy the entire gap $V_f/V_{g0} = 0.9$. Dashed curves are presented for comparison with the cases of purely elastic solids, discussed in Sec.~\ref{sec::results_incomp}. Vertical dash-dotted line indicates the hardness taken to be $H = 3\sigma_Y$.}
\label{fig::comp_nonlin_incomp_el_pl}
\end{figure}

As was shown in Fig.~\ref{fig::comp_lin}(a) for the case of linearly compressible fluid (with constant bulk modulus), starting from the pressurization of the fluid, the real contact area evolves monotonically with the external pressure: if the fluid bulk modulus is less than the one of the solid ($K_f < K_s$), then the real contact area increases up to the full contact state, if $K_f > K_S$, then the contact area decreases down to zero, corresponding to the opening of the trap. The latter case is interesting for the study of the process of the fluid permeation into the contact zone and reduction of the global coefficient of friction, however, as it was mentioned in the Sec.~\ref{sec::results_incomp} for the incompressible fluid, the situation when the initial fluid bulk modulus is greater than that of the solid remains non-physical and serves as an idealized model.
On the other hand, real fluids behave non-linearly and their bulk modulus increases with increasing pressure, and thus even if the fluid bulk modulus is smaller than that of the solid in the first stage of pressurization, it eventually becomes greater than the one of the solid under the increasing pressure.

We present results of the numerical simulation for coupled problem with non-linear fluids - evolution of the contact area and global coefficient of friction with respect to increasing external pressure, see Figs.~\ref{fig::comp_nonlin_incomp_el_pl}(a),(c), respectively. Physically relevant values for two solid materials are used: a typical steel ($E = 200 \text{ GPa}, \nu = 0.28, K_s \approx 151.5 \text{ GPa}$) and aluminium ($E = 70 \text{ GPa}, \nu = 0.33, K_s \approx 83.33 \text{ GPa}$), and three types of fluid (see Eq.~\eqref{eq::comp_modulus_nonlin}): water ($K_0 = 2112.5 \text{ MPa}, K_1 = 6.5$), glycerine ($K_0 = 4151.5 \text{ MPa}, K_1 = 8.74$) and a typical mineral oil ($K_0 = 2000.0 \text{ MPa}, K_1 = 9.25$)~\cite{Kuznetsov_1985,Nellemann_1977}.
We limit this study to the contact problem with the fluid completely filling up the gap (but only up to the upper boundary) during the whole process of loading. Such formulation remains rather general since, due to the realistic fluid model, the contact zone will inevitably appear under the first loading.

At low external pressures numerical results coincide with the analytical solutions also obtained for non-linear fluids, see~\eqref{eq::p_kuz_nonlin}. However, in contrast to the analytical solution, which cannot account for depletion of the contact zone, the numerically obtained contact area, as expected, becomes a non-monotonic function of pressure and after reaching the maximum, decreases 
Note that for each of considered materials, the obtained curves for water and oil coincide in the beginning of loading due to almost equal initial bulk moduli $K_0$ of these fluids, and deviate for higher external pressures due to difference in $K_1$, while for glycerine $K_0$ is significantly bigger, leading to a smaller contact area in this case.

The global coefficient of friction (CoF) also shows a non-monotonic behaviour, see Figs.~\ref{fig::comp_nonlin_incomp_el_pl}(c), it vanishes when the contact area is zero, and rapidly increases up to a certain maximal value. Within this stage, the numerical and analytical results are very close, while for higher pressures a strong deviation of analytical and numerical results is observed. In analytical solution, even though the global CoF may decrease after the first extremum-maximum (see results obtained for the steel), it eventually increases again after reaching the second extremum-minimum. More accurate numerical results predict a monotonic decrease of the global CoF after reaching the first maximum. Note that in the simplified case considered here, the hydrostatic lubrication effect decreases significantly the maximal global CoF, which does not exceed $\approx$ 36 \% of the local CoF for the steel, and does not exceed $\approx$ 24 \% of the local CoF for the aluminium. Such a strongly non-linear behaviour of the global coefficient of friction (with one or two extrema) is explained by a competition between non-linear fluid pressurization and non-linear contact area evolution (see Eq.~\eqref{eq::est3}).

The numerical solution shows that the maximal value of the CoF and its slope after passing the extremum both depend on the ratio between the bulk moduli of the fluid $K_f = K_0 + K_1 p_f$ and the solid $K_s$. The bigger is the initial modulus $K_0$, the higher is the maximal CoF 
(which explains almost equal peak values of the CoF for water and oil and much lower value for glycerine). 
At the same time, the bigger is the coefficient $K_1$, the faster the CoF decreases.

We performed additional simulations varying the slope of the roughness profile $\Delta/\lambda$ in the interval $[0.005; 0.02]$. The results showed that the evolution of the real contact area is almost independent of the ratio $\Delta/\lambda$ (similarly to the case of the incompressible fluid). 
On the other hand, variation of this ratio has a considerable effect on the peak value of the global CoF, which increases with increasing $\Delta/\lambda$. 
However, for high values of external pressure, the CoF does not depend on the slope of the profile, as it was also observed for the incompressible case.

It is important to note here that, even if in the numerical solution the contact area decreases with the increasing external pressure, it does not reach zero value even for extremely high values of the external pressure $p_0=E^*$. Thus using linearly elastic material model seems to be irrelevant at such high pressures. In Section~\ref{sec::elasto_plastic} a more realistic case will be presented taking into account a non-linearly compressible fluid and a relevant elasto-plastic material behaviour.

\subsection{Elastic-perfectly plastic solid\label{sec::elasto_plastic}}

\begin{figure}[h]
	\centering
	\includegraphics[width=0.99\textwidth]{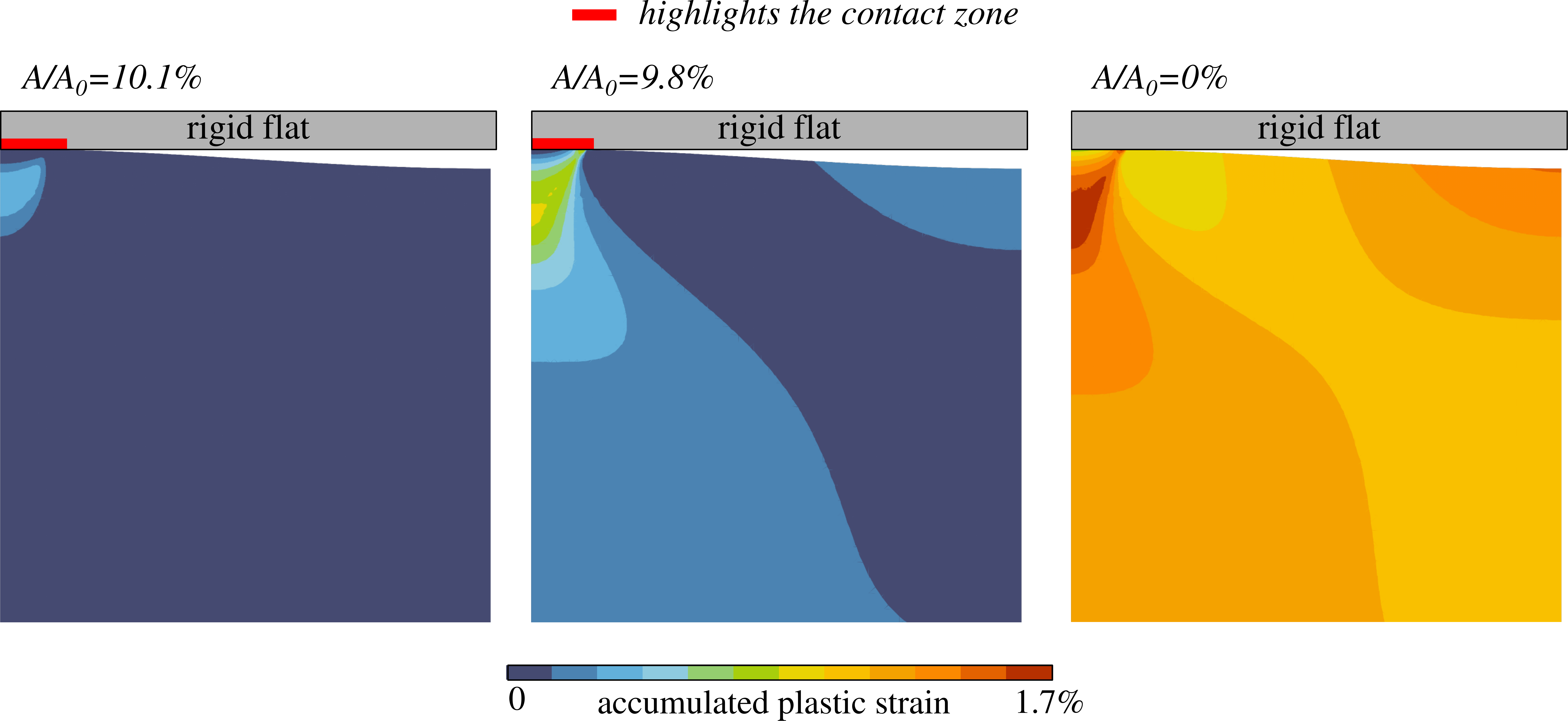} 
	\caption{Accumulated plastic strain near the contact interface is shown at three different external loads for the incompressible fluid and for $V_f/V_{g0}=0.9$ and $\Delta/\lambda= 0.01$, from left to right: (1) the step corresponding to activation of the fluid (maximal contact area); (2) contact area decreased by half; (3) zero contact area (the trap is opened; as seen from the plastic field, at this moment the entire solid is plastified).}
	\label{fig::el_pl_map}	
\end{figure}

Here we consider elastic-perfectly plastic materials (von Mises stress criterion): steel, $E = 200 \text{ GPa}$, $\nu = 0.28$, yield stress $\sigma_Y = 250 \text{ MPa}$ and aluminium, $E = 70 \text{ GPa}, \nu = 0.33, \sigma_Y = 240 \text{ MPa}$. 

It is well known that in elasto-plastic mechanical contact, the contact pressure cannot exceed the material hardness, which can be reliably estimated as $H\approx3\sigma_Y$~\cite{Bowden_2001,Johnson_1985,Mesarovic1999spherical}. Thus it could be expected that after the pressure in the fluid reaches material hardness the contact abruptly opens. However, as demonstrated by our simulations, due to the high hydrostatic compressive state, the pressure in the contact can significantly overpass the material hardness. 

First, we study incompressible fluid, and present in Fig.~\ref{fig::comp_nonlin_incomp_el_pl}(b) the evolution of contact area in the case of $V_f / V_{g0} = 0.9$. It shows significantly different behaviour compared to elastic material: after the fluid becomes activated, the contact area is non-monotonic function of external pressure, it has a small increase, and then an abrupt decrease, corresponding to the state when fluid pressure reaches the value of contact pressure, and, consequently, permeation becomes possible. Normal tractions in contact interface increase beyond $6 \sigma_Y$ - due to hydrostatic pressurization of the solid. In Fig.~\ref{fig::comp_nonlin_incomp_el_pl}(d) the resulting evolution of the global CoF is presented, which shows considerably lower values of the CoF for the both considered materials, than the ones observed in the purely elastic case (for the same external pressure). Fields of the accumulated plastic strain in the solid at different loading steps are presented in Fig.~\ref{fig::el_pl_map}, note that once the fluid gets pressurized, the plastic zone is not limited to the contact vicinity, but spreads over the entire interface and, consequently, the whole bulk of the solid. Notably, a secondary onset of plastic deformation appears in the trough of the wavy profile, it complements the classical plastic core appearing under the contact zone and spreading to the contact interface~\cite{Johnson_1985,Mesarovic1999spherical,kogut2002elastic,alcala2010reassessing}.

Varying the slope of the profile as in Sections~\ref{sec::results_incomp} and~\ref{sec:results_comp_nonlin}, we showed that in contrast to the case of elastic solids, where the evolution of the contact area during the process of trap opening does not depend on the slope of the profile $\Delta/\lambda$, in case of elasto-plastic solids, for a given ratio $V_f / V_{g0}$, once the fluid gets pressurized, the higher is the ratio $\Delta/\lambda$, the bigger is the contact area.

\begin{figure}[p]
	\centering
	\includegraphics[width=0.99\textwidth]{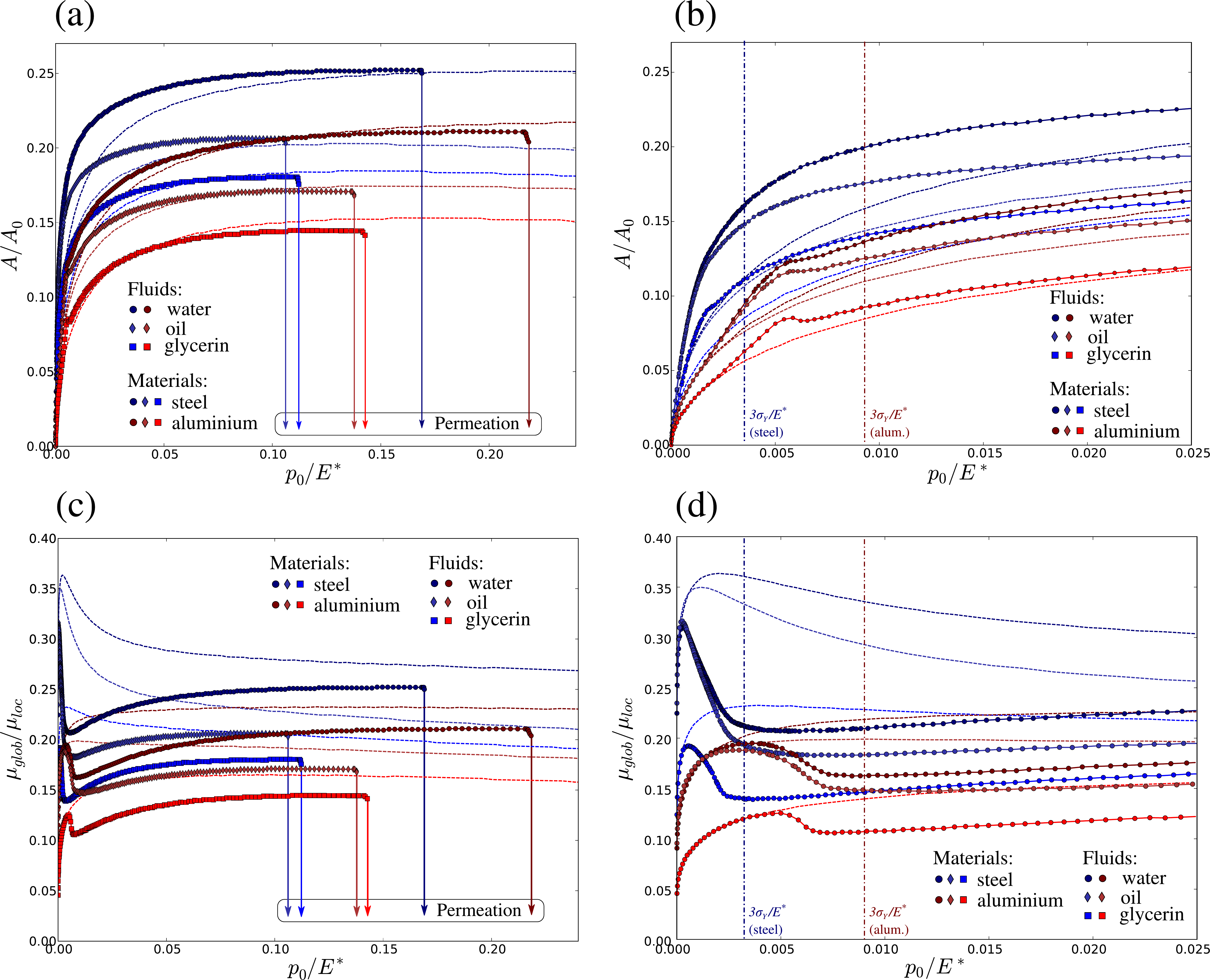} 	
	\caption{The behaviour of the system considering elasto-plastic material and non-linearly compressible fluid: (a) evolution of the ratio of real contact area to the apparent one under increasing external pressure; (b) the same as (a), but the results are shown in range $0 \leq p_0 \leq 0.025 E^*$; (c) evolution of the ratio between global and local coefficients of friction; (d) the same as (c), but the results are shown in range $0 \leq p_0 \leq 0.025 E^*$. Dashed curves are presented for comparison with the cases of purely elastic solids. Vertical dash-dotted line indicates the hardness $p_0 = H = 3\sigma_Y$.}
	\label{fig::el_pl}
\end{figure}

The behaviour of the system incorporating the elasto-plastic material and non-linearly compressible fluid is shown in Figs.~\ref{fig::el_pl}(a-d): the contact area after reaching its maximum abruptly decreases, resulting in a fast permeation of the fluid in the contact interface and eventual opening of the contact. Note that after a relatively fast saturation of the contact pressure at approximate material hardness $H\approx3\sigma_Y$, a further increase in pressure without fluid permeation still remains possible up to huge pressure values $p_0 \gg \sigma_y$. In reality however, due to the micro-roughness permeation of the fluid in the contact interface may happen on earlier stages of the deformation.

In fig.~\ref{fig::el_pl}(c,d) the evolution of the global CoF is depicted, which shows a rather similar behaviour to the one observed in the case of the elastic solid, having multiple extrema in the beginning of loading. Note that the amplitude of the first maximum of CoF is increasing with increasing slope of the profile, which was also observed in the simulations with the purely elastic material.

\subsection{Friction in the contact interface}

\begin{figure}[p]
\centering
\includegraphics[width=0.99\textwidth]{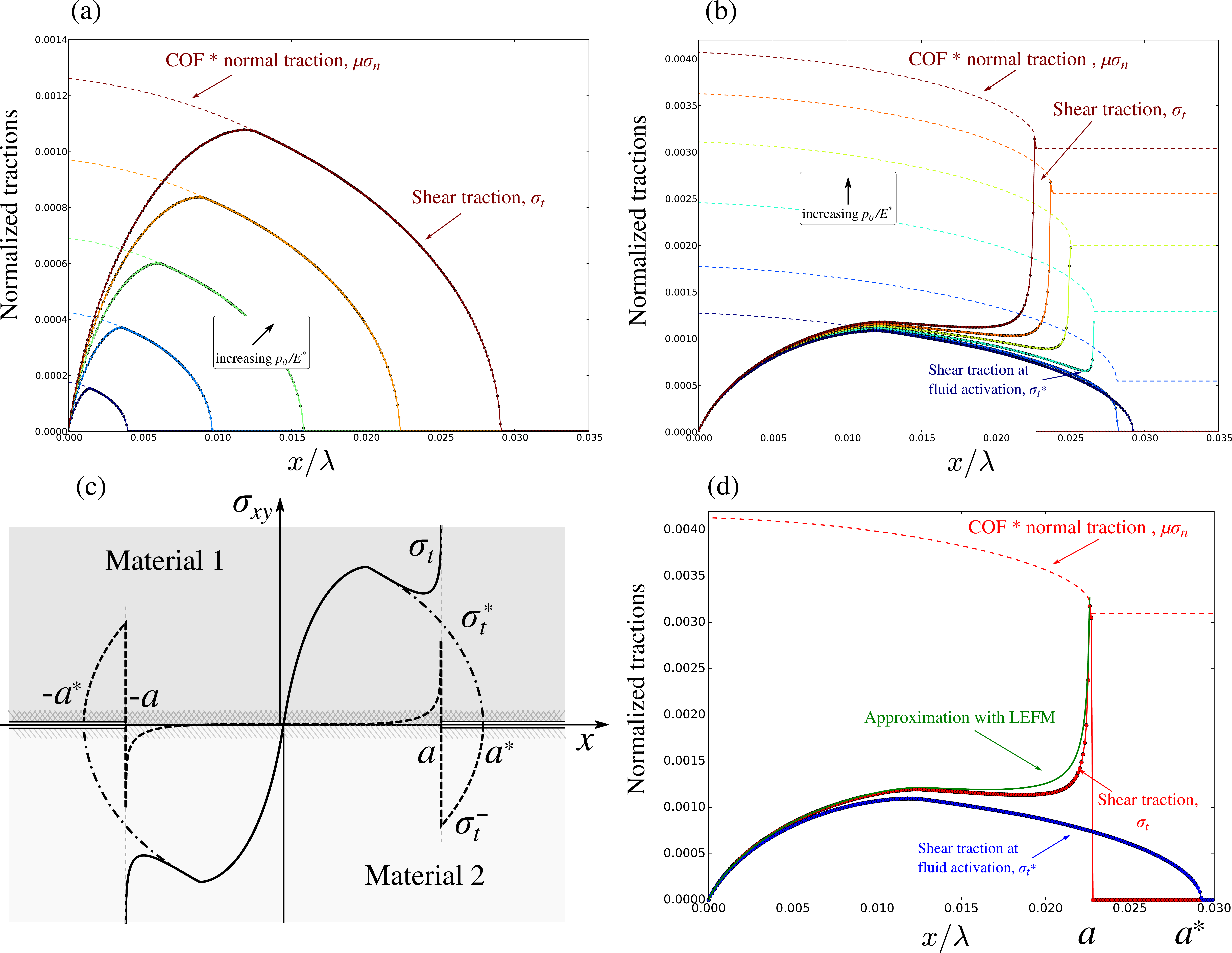} 
\caption{Distribution of the tangential tractions in the contact interface: (a) Fluid is not pressurized. (b) Under increasing external load fluid gets pressurized, contact area is decreasing and a singularity in tangential traction appears (limited by the Coulomb's law). (c) Sketch of the analogous problem 
for two bonded dissimilar solids with two aligned semi-infinite interfacial cracks in the interface. (d) Comparison of the numerical results for the shear tractions and approximation provided by the analogy with the LEFM.}
\label{fig::incomp_fric}
\end{figure}

In order to study the distribution of frictional tractions in the contact interface during the process of opening of the trap, we consider a coupled problem for an incompressible fluid with Coulomb's friction in the contact interface, as in previous analysis the shear forces in the trapped fluid are neglected because of quasi-static analysis.
The following geometrical parameters are used: $\Delta/\lambda = 0.01$, $V_f/ V_{g0} = 0.95$. In order to obtain more reliable results, we refined the mesh to have 512 nodes within the maximal extension of the contact zone $a/\lambda = 0.05$, with 1024 surface elements in total.

Two stages in the loading can be distinguished. During the first stage the external pressure $p_0$ increases from zero value to $p_{\text{act}}$, the value necessary to bring the fluid into active state, and the contact area reaches the maximum value. Results for the first stage are presented in Fig.~\ref{fig::incomp_fric}(a), where, in order to visualize stick and slip zones, we plot normal tractions, multiplied by the coefficient of friction (CoF) $\mu = 0.2$.
Those results are very close to the classic self-similar (remaining the same for any load under a proper coordinate/pressure scaling~\cite{Spence_1968}) distribution of tractions, because the wavy profile in the region of interest is very close to a parabolic curve.
During the second stage of loading ($p_0 > p_{\text{act}}$) the fluid is in the pressurized state and influences the interfacial traction distribution.

Since the slope of the roughness profile is small, the distribution of normal traction should resemble, at least for $p_0$ not much greater than $p_{act}$, the analytical solution for a fluid bulk modulus tending to infinity ($K\to\infty$), in which a uniform pressure offset is added everywhere to the field of the normal traction corresponding to the external pressure $p_{act}$. In accordance to that, tangential traction remains almost unchanged over the majority of the contact interface. Because of the contact pressure increase by the fluid pressure offset, all points pass to the stick state, i.e. adhere to their positions. However, due to the finiteness of the slope being taken into account, the distribution of normal traction slightly differs from the analytical solution in the same way as was discussed in Sec.~\ref{sec::results_comp_lin}, see Fig.~\ref{fig::comp_lin}(b), i.e. a slight decrease of the contact area takes place. 

For $p_0$ sufficiently greater than $p_{\text{act}}$, see Fig.~\ref{fig::incomp_fric}(b), the effects of finite slope become more pronounced, the contact area is gradually decreasing and a remarkable evolution of the tangential traction is observed. A singularity in the tangential traction emerges at the boundary of the contact zone, with the value at the tip of this singularity limited by the Coulomb's law. In order to explain and verify this intriguing result, we consider an analogy between the process of the trap opening with the interfacial friction and the mode-II crack propagation in the framework of linear elastic fracture mechanics (LEFM) theory~\cite{tada1973stress}.

Note that the analogy is not complete in physical sense: during the process of trap opening due to pressurization of the incompressible fluid, new surface is not created, since no atomic bonds must be broken in order to separate the surfaces. The physical reason for the singularity in tangential stress is the following: when points of the surface loose contact, their normal traction reduces not down to zero, but to the value of fluid pressure, thus the frictional limit near the contact edge remains elevated. Thus, the points of the interface before loosing contact have non-zero shear traction, and being liberated from this traction after loosing the contact, these points slide freely, in absence of frictional resistance, towards the centre of the contact zone. 

The fluid activation corresponds to the maximal extension of the contact zone (we shall denote the maximal contact half-length as $a^*$, and during the subsequent increase of the external pressure the width of the contact zone is monotonically decreasing. For sufficiently small slope of the roughness profile, the situation corresponding to contact half-length $a < a^*$ can be considered as a configuration of two bonded dissimilar solids with two aligned semi-infinite interfacial cracks in the interface, separated by $2a$, see Fig.~\ref{fig::incomp_fric}(c). Using the superposition principle, the observed stress state, corresponding to the half-length of the contact patch $a$, can be represented as a superposition of the initial shear traction $\sigma_{t}^*(x)$, corresponding to the moment of activation of the fluid, and a stress induced by the same traction with the opposite
sign, $\sigma_t^-(x) = -\sigma_{t}^*(x)$ applied only on the surfaces of the cracks in the intervals $x\in[-a^*, -a]\:\text{and}\:[a, a^*]$.
Such traction induces a singular shear stresses in the region between two cracks $x \in[-a, a]$, thus $\sigma_t^-(x)$ can be written as:
\begin{equation}
\sigma_t^-(x) = 
\begin{cases}
-\sigma_{t}^*(x),& \; x \in[-a^*, -a] \cup [a, a^*]\\
\frac{1}{\sqrt{2\pi}}\text{Im}\left\{K(a, \sigma_t^*)\ \left(\frac{(x-a)^{i \epsilon}}{\sqrt{|x-a|}} - \frac{(x+a)^{i \epsilon}}{\sqrt{|x+a|}}\right)\right\},& \; x \in[-a, a]\\
0,& \; |x| > a^*,
\end{cases}
\label{eq::lefm}
\end{equation}
where $K$ is the complex stress intensity factor, see ~\cite{rice1965plane,rice1988elastic}, and two terms in brackets in~(\ref{eq::lefm}.2) correspond to two semi-infinite cracks being considered, so that $\sigma_t^-(0) = \sigma_t^*(0) = 0$, $\text{Im}$ is the imaginary part. 
Therefore, the resulting distribution of shear tractions is given by the superposition $\sigma_t(x) = \sigma_t^*(x) + \sigma_t^-(x)$. 

The complex stress intensity factor $K$ is calculated using the existing analytical formula for considered configuration and shear traction distribution~\cite{rice1965plane,rice1988elastic}: 
\begin{equation}
K(a, \sigma_t^*) = \left[k_1(a, \sigma_t^*) + i k_2(a, \sigma_t^*)\right] \sqrt{\pi} \cosh{(\pi \epsilon)},
\label{eq::lefm_k}
\end{equation}
where
\begin{align}
&k_1(a, \sigma_t^*) = \frac{\sqrt{2}}{\pi}\int\limits_a^{a^*}  \frac{\sigma_t^*(x) \sin{(\epsilon \ln{(x-a)})}}{\sqrt{x-a}} dx, \nonumber \\
&k_2(a, \sigma_t^*) = \frac{\sqrt{2}}{\pi}\int\limits_a^{a^*}  \frac{\sigma_t^*(x) \cos{(\epsilon \ln{(x-a)})}}{\sqrt{x-a}} dx,
\label{eq::lefm_k1_k2}
\end{align}
and the parameter $\epsilon$ accounts for the different properties of the two bonded solids, in case one of them being rigid, it equals to
\begin{equation}
\epsilon = -\frac{1}{2\pi} \ln{(3 - 4 \nu)}.
\end{equation}
In Fig.~\ref{fig::incomp_fric}(d) we plot the approximation of the shear traction distribution in the interface during trap opening, discussed above. 
A sound similarity is found between numerical results and analytical formulae provided by the LEFM.
Therefore, we have shown that during the process of trap opening due to increasing pressure in the fluid with friction taken into account, the tangential tractions near the contact edges are elevated up to the limit provided by the Coulomb friction law. Consequently, even if the majority of the interface remains in stick state, local slip zones emerge at the boundaries of contact zones. It is important to account for such an elevated shear stress near edges of contact zones, which surround trapped fluid, in the analysis of damage evolution and crack onset under monotonic and cycling loading, including fretting fatigue \cite{hills1994mechanics,proudhon2005fretting}.

\section{Conclusions}
\label{sec::conclusions}

In this work we solved the problem of mechanical contact between a deformable body with a wavy surface and a rigid flat, taking into account pressurized fluid trapped in the interface. 
A mathematical framework for this coupled problem for both incompressible and compressible fluids was formulated. 
In the latter case, either constant or pressure-dependent fluid bulk-moduli were considered; all models were implemented in the finite element framework using a monolithic approach.

The proposed framework accounts for a finite slope of the roughness profile, while in previous investigations using classical boundary element method (which accounts only for vertical displacements) and existing analytical solutions only infinitesimal slopes were considered. 
We show that in the considered coupled problem, a reduction of the contact area can occur due to elastic flattening of asperities by fluid pressure. 
Thus the reduction of the global coefficient of friction is caused not only by the external load repartition between the solid contact and the pressurized fluid, but also by the contact area reduction.

The reduction of the contact area takes place if the fluid bulk-modulus is higher than that of the solid. 
In case of incompressible fluid this criterion is satisfied and the process of trap opening is observed. 
However, this case is non-physical, since real lubricating fluids in the unpressurized state have much lower bulk modulus than solids. 
A more relevant case is a compressible fluid with linear dependence of bulk modulus on pressure, which ensures a non-monotonic variation of the contact area, and thus of the global coefficient of friction, leading to reduction of the both for sufficiently large pressures.

Among other applications, the obtained results are relevant for the mechanical behaviour of multi-cracked materials such as rocks. We showed that due to the presence of pressurized fluid in the interface, the frictional limit does not increase linearly with increasing external load, but reaches its maximum and decreases down to zero. 
This behaviour is similar to pressure-dependent plasticity models with a cap (e.g. Drucker-Prager cap model), which corresponds to the decay of the von Mises yield stress with the increasing pressure.

In addition to elasticity, we considered physically more relevant elasto-plastic materials in combination with realistic fluids.
In this case, the contact pressure is bounded, while the fluid can bear arbitrary pressure, consequently under certain external pressure fluid permeates in the contact zones abruptly.

When interfacial friction is considered in the coupled problem, previously unreported quasi-singularities appear in shear stresses near edges of contact patches during fluid-trap opening.
We showed that these singularities can be analytically estimated using the analogy between trap opening and crack propagation in the interface between two bonded dissimilar solids.
It is important to account for such an elevated shear stress, caused by the trapped fluid, in the analysis of damage evolution and crack onset under monotonic and cycling loading, including fretting fatigue.

The problem of trapped fluid is relevant for metal forming (drawing and rolling), where a lubricant is present in the interface and involved loads are high.
It is also relevant in poromechanics, especially in cracked media filled with fluid and subjected to complex stress states with high hydrostatic component, which can ensure contact between surfaces of internal cracks. 
Finally, at the microscopic scale, where the surface roughness plays a crucial role, the trapped fluid provides additional load-bearing capacity, and thus reduces the macroscopic static friction. 
Under increasing load, the trapped fluid is squeezed out of its trap thus resulting in even smaller global coefficient of friction.

\section{Acknowledgements}
The authors acknowledge the financial support of Safran Tech and MINES ParisTech (Th\`ese-Open) and are grateful to Julien Vignollet for his helpful suggestions and kind support. 
Enlightening comments of an anonymous reviewer are kindly acknowledged.

\appendix

\section{
Distortion of the periodic wavy surface under uniform normal pressure\label{app:aux}}

 \begin{figure}[h]
	\begin{center}		
		\includegraphics[width=0.5\textwidth]{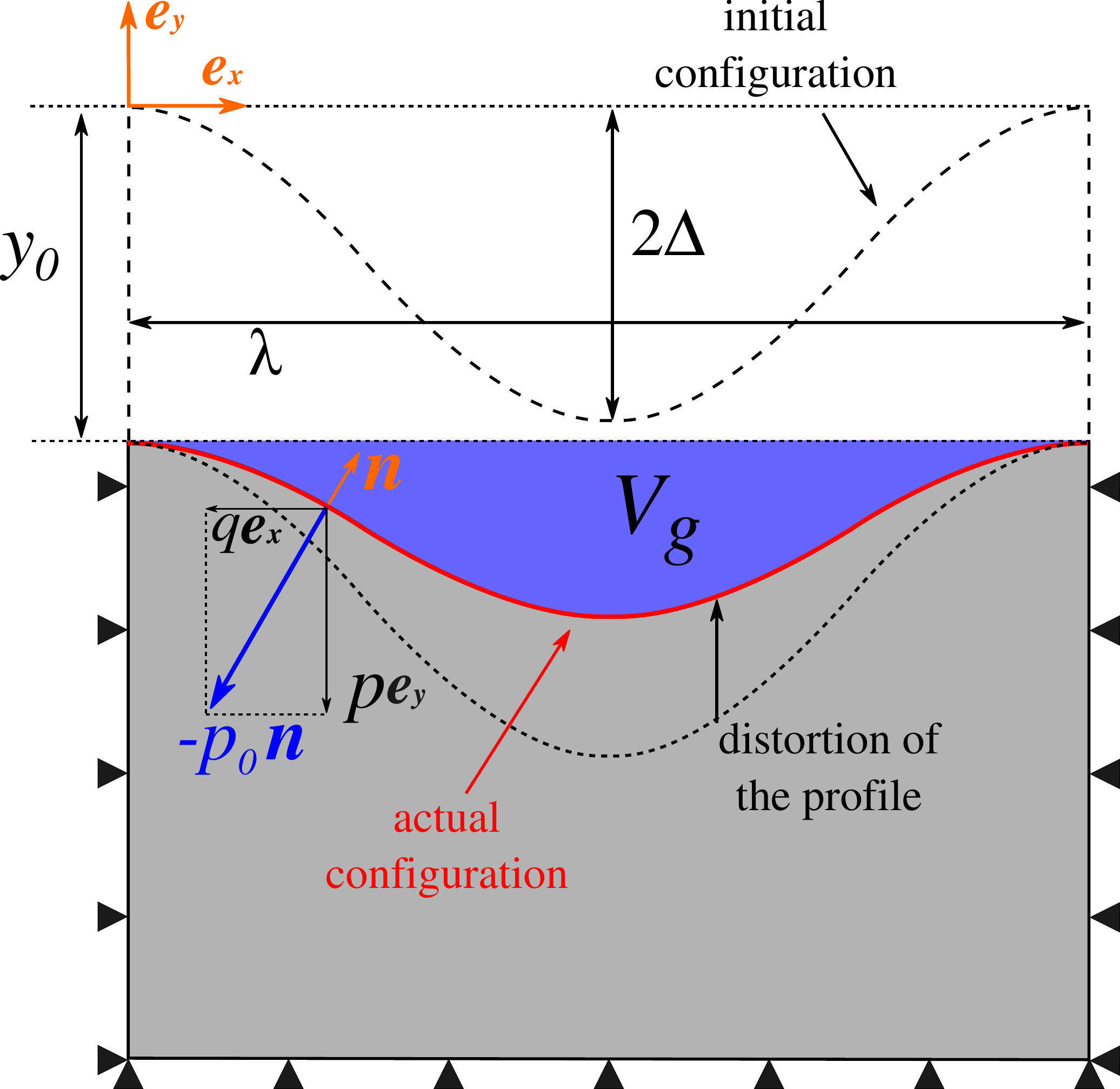} 
		\caption{
Distortion of a periodic wavy surface under a uniform pressure.}
		\label{fig::aux_appendix}
	\end{center}
\end{figure}

According to the integrated Flamant's solution, any uniformly distributed pressure on the surface will result in a uniform vertical displacement. However, it is true only if the surface is flat. For a wavy surface, under the action of a uniform pressure, the crest's displacement is bigger than the displacement of the trough. The uniform pressure distribution on the surface is given by $-p_0\vec{n}$, where $\vec{n}$ is the outer normal to the surface (see Fig.~\ref{fig::aux_appendix}). 
We consider the vertical $p$ and horizontal $q$ components of the normal pressure (each one of them contributes to the distortion of the profile): 
$$q(x) = -2\pi p_0 \frac{\Delta}{\lambda}\sin \frac{2\pi x}{\lambda} + O\left(\frac{\Delta^3}{\lambda^3}\right)$$ and $$p(x) = -p_0 + O\left(\frac{\Delta^2}{\lambda^2}\right)$$  in case of small slope. 
Therefore, keeping the small values of order $\Delta/\lambda$, we may calculate the vertical displacement caused by the horizontal component $q$ using the integrated Flamant's solution:
$$u_y^q(x)=-\frac{(1-2\nu)(1+\nu)}{2E}\left\{\int\limits_{-b}^{x}q(s)\:ds - \int\limits_{x}^{b}q(s)\:ds\right\}+C,$$
where $b\rightarrow\infty$ and $C$ is an arbitrary constant. Substituting $q(s)$ and calculating integrals, we obtain:
\begin{equation}
u_y^q(x)=-\frac{(1-2\nu)(1+\nu)p_0}{E}\Delta\cos\frac{2\pi x}{\lambda}+C.
\label{eq:q}
\end{equation}
We are also convinced that the uniformly distributed vertical traction $p$ produces exactly the same vertical displacement:
\begin{equation}
u_y^p(x)=-\frac{(1-2\nu)(1+\nu)p_0}{E}\Delta\cos\frac{2\pi x}{\lambda}+D,
\label{eq:p}
\end{equation}
where $D$ is another arbitrary constant. However, this result does not follow from Flamant's solution, which as already mentioned would predict a uniform displacement. This result was guessed and confirmed with a very high accuracy (fractions of percent) by finite-element simulations for different fractions $\Delta/\lambda$ and Poisson's ratios. Of course, the simplest analogy would be the Winkler's foundation with springs whose lengths follow the distribution $l(x) = L + \Delta\cos(2\pi x/\lambda)$, which would mimic the shape of the wavy surface. However, it is unavailing to obtain the proportionality factor of form $(1-2\nu)(1+\nu)/E$.  At this stage we are content with numerical proof only, which consisted in applying separately horizontal and normal components of the pressure over a single period with periodic boundary conditions and comparing the numerical results with the equations~\eqref{eq:p} and \eqref{eq:q}.
The total displacement field reads:
$$u_y(x) = u_y^q(x) + u_y^p(x) = -\frac{2(1-2\nu)(1+\nu)p_0}{E}\Delta\cos\frac{2\pi x}{\lambda}+\tilde{C}.$$
We may define $\tilde{C}$ so that $u_y(0) = y_0$, where $y_0$ is the current vertical position of the crest, thus giving:
$$u_y(x) = y_0-\frac{2(1-2\nu)(1+\nu)p_0}{E}\Delta\left(\cos\frac{2\pi x}{\lambda}-1\right).$$
We calculate the volume between the distorted wavy surface and the plane $y=y_0$:
$$V_g = \int\limits_{\Gamma} (y_0 - (Y(x) + u_y(x))) d\Gamma,$$
where $Y(x) = \Delta(\cos\frac{2\pi x}{\lambda}-1)$ is the initial vertical coordinate. Note that horizontal displacements do not contribute significantly to the volume change. Upon integration, noting that the initial gap volume $V_{g0} = \Delta \lambda$, we obtain:
$$V_g/V_{g0} = 1 - \frac{2(1-2\nu)(1+\nu)p_0}{E},$$
which gives the relative change of the volume between the deformed wavy surface and corresponding plane under action of the uniform normal pressure $p_0$. Note that it does not depend on $\Delta/\lambda$, however, this result was obtained under assumption of $\Delta/\lambda\ll1$, therefore it corresponds to the limiting case of small, but finite slope of the profile.

\section{Trapped fluid element}
\label{sec::appendix}
\begin{figure}[t]
\begin{center}		
	\includegraphics[width=0.4\textwidth]{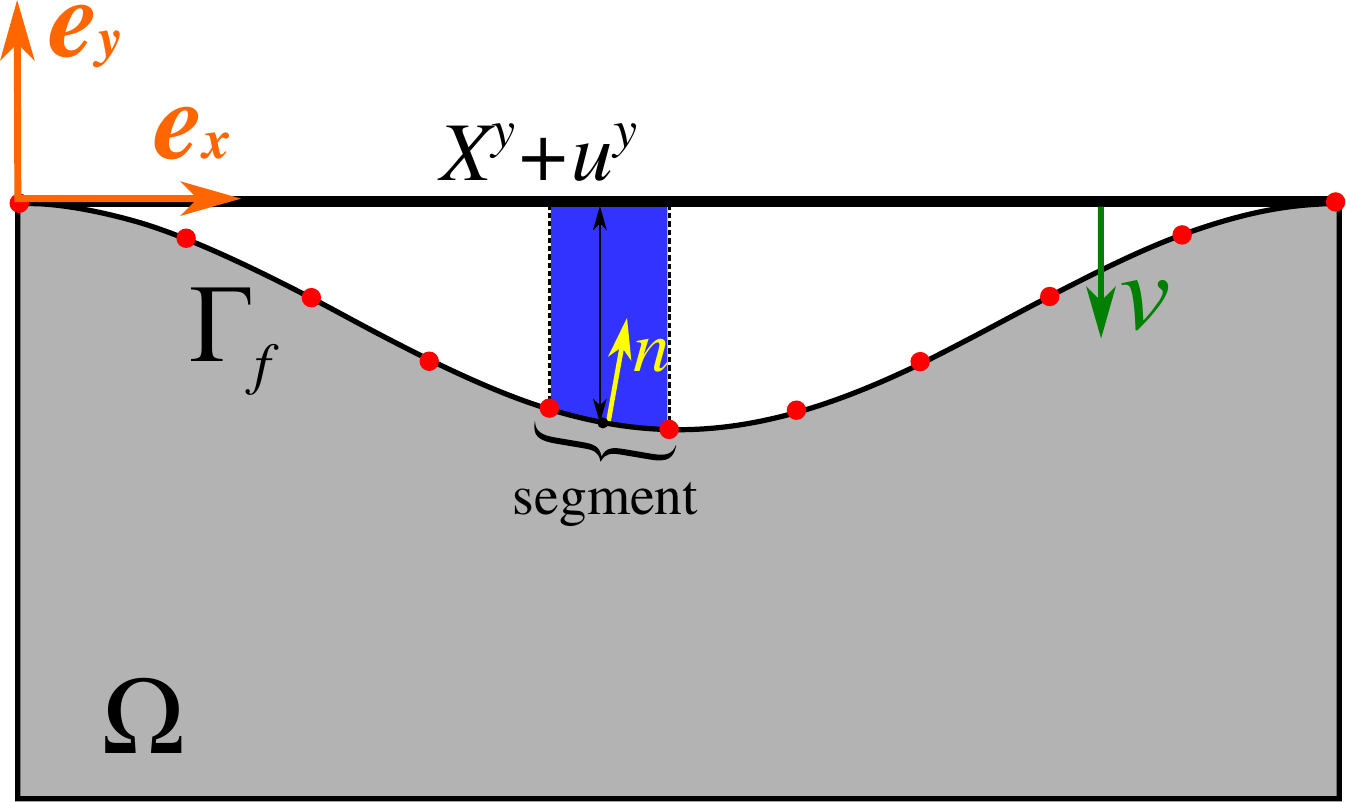} 
	\caption{Trapped fluid element, consisting of all segments of the surface boundary $\Gamma_f$ (shaded in blue is the part of the trapped fluid volume, which corresponds to the highlighted segment).}
	\label{fig::element}
\end{center}
\end{figure}

In order to implement in a finite element code our approach for modelling the trapped fluid, alongside with structural and contact elements, we used a special \textit{trapped-fluid element} containing all segments of the trapped fluid zone $\Gamma_f$, see Fig.~\ref{fig::element} (in the FEM literature this element is also known as a hydrostatic fluid element). 

In the finite element framework the area of the gap~(\ref{eq::gap_volume_integral}) can be calculated by the following formula:
\begin{equation}
\label{eq::gap_volume_elem}
V_g = \sum_{\text{seg}} \int_{-1}^{1} (X^y_i + u^y_i)N_i(\xi) n^y_{\text{seg}}(\xi) J(\xi) \: d\xi,	
\end{equation}
where the summation is performed over all segments of the surface $\Gamma_f$, 
$X^y_i$ and $u^y_i$ are the vertical coordinate in the reference configuration and the vertical displacement of the $i$-th node of the corresponding segment, respectively,
$N_i(\xi)$ is the shape function, associated with the $i$-th node;
$n^y_{\text{seg}}(\xi) = \boldsymbol \nu \cdot \boldsymbol n_{\text{seg}}(\xi)$, where
$\boldsymbol{\nu}$ is the normal to the rigid plane, and $\boldsymbol n_{\text{seg}}(\xi)$ is the normal to the segment, 
$J(\xi)$ is the Jacobian and $\xi\in[-1,1]$  is the convective coordinate in \textit{the parent} space. Note that summation over the repeating indices is assumed.

Therefore we can consider the gap volume~(\ref{eq::gap_volume_elem}) as a function of the displacement vector $\mathbf{u} = [u_1^x,u_1^y,\dots u_N^x,u_N^y]^T$, consisting of the displacement components of $N$ nodes on the surface $\Gamma_f$, $(u^x_i, u^y_i)$ are horizontal and vertical components of the displacement vector of the $i$-th node, respectively.

\subsection{Lagrange multiplier formulation}

The problem of finding a stationary point of the Lagrangian~(\ref{eq::Lagrangian_coupled}) is non-linear, and to solve it numerically we use the classical Newton-Raphson method, which requires calculation of the residual vector of the trapped-fluid element
\begin{equation}
\label{eq::res_vector_tf}
[\textbf{R}_{f}] = 
\begin{bmatrix}
\lambda_f\left[\frac{\partial V_g(\mathbf{u})}{\partial\mathbf{u}}\right] \\[10pt]
V_g(\mathbf{u}) - V_f
\end{bmatrix}_{(2N + 1) \times 1}
\end{equation}
and the corresponding tangent matrix
\begin{equation}
\label{eq::tang_matrix_tf}
[\textbf{K}_{f}]= 
\begin{bmatrix}
\lambda_f\left[\frac{\partial^2 V_g(\mathbf{u})}{\partial\mathbf{u}^2}\right] && \left[\frac{\partial V_g(\mathbf{u})}{\partial\mathbf{u}}\right] \\[10pt]
\left[\frac{\partial V_g(\mathbf{u})}{\partial\mathbf{u}}\right]^T && 0
\end{bmatrix}_{(2N + 1)\times (2N + 1)}.
\end{equation}
Note that the vector of degrees of freedom (DOFs) for the trapped fluid element has the form $[u_1^x,u_1^y,\dots u_N^x,u_N^y,\,\lambda_f]^T$, and includes  $2N$ displacement DOFs of $N$ surface nodes and also an additional single Lagrange multiplier $\lambda_f$ (which represents the fluid pressure).

On every global iteration of the Newton-Raphson method we perform the following \textit{active set strategy}, see~\cite{Wriggers_2006,Yastrebov_2013} for the trapped fluid element:
\begin{enumerate} 
\item Get from previous iteration current value of the displacement vector $\mathbf{u}$ and the Lagrange multiplier $\lambda_f$, calculate the gap volume $V_g(\mathbf{u})$;	
\item If the gap $V_g(\mathbf{u}) > V_f$ or $\lambda_f < 0$, the fluid is considered to be in \textbf{inactive} state and the trapped fluid element is excluded from consequent calculations;
\item Otherwise, i.e. if $V_g(\mathbf{u}) \leq V_f$ and $\lambda_f \geq 0$, the fluid is in \textbf{active} state, then the residual vector and the tangent matrix are calculated by formulas~(\ref{eq::res_vector_tf}) and~(\ref{eq::tang_matrix_tf}).
\end{enumerate}

\subsection{Penalty formulation}

For the numerical simulations of the trapped compressible fluid we may consider the same finite element, as was described previously for the case of Lagrange multipliers, with one difference: no extra degrees of freedom are involved, and the vector of DOFs has the form $[u_1^x,u_1^y,\dots u_N^x,u_N^y]^T$. A similar active set strategy can be applied to this element, as it was described for the case of Lagrange multiplier element. 

The residual vector and the tangent matrix for the linear penalty trapped fluid element have the form:
\begin{equation}
\label{eq::res_vector_tf_penalty}
[\textbf{R}_{f}] = \frac{K}{V_{f0}}
\begin{bmatrix}
\left(V_g(\mathbf{u}) - V_{f0}\right)  \: \frac{\partial V_g(\mathbf{u})}{\partial \mathbf{u}}
\end{bmatrix}_{2N\times 1},
\end{equation}
\begin{equation}
\label{eq::tang_matrix_tf_penalty}
[\textbf{K}_{f}] = \frac{K}{V_{f0}}
\begin{bmatrix}
\frac{\partial V_g(\mathbf{u})}{\partial \mathbf{u}} \otimes \frac{\partial V_g(\mathbf{u})}{\partial \mathbf{u}} + \left(V_g(\mathbf{u}) - V_{f0}\right)  \: \frac{\partial^2 V_g(\mathbf{u})}{\partial \mathbf{u}^2}
\end{bmatrix}_{2N\times 2N},
\end{equation}
where $\otimes$ is a tensor product, see~\cite{Yastrebov_2013}. In the case of a non-linear penalty element:
\begin{equation}
\label{eq::res_vector_tf_nonlin_penalty}
[\textbf{R}_{f}] =
\begin{bmatrix}
-\frac{K_0}{K_1}\left\{\left(\frac{V_g(\mathbf{u})}{V_{f0}}\right)^{-K_1}-1\right\}\frac{\partial V(\mathbf{u})}{\partial\mathbf{u}}
\end{bmatrix}_{2N\times 1},
\end{equation}
\begin{equation}
\label{eq::tang_matrix_tf_nonlin_penalty}
[\textbf{K}_{f}] =
\begin{bmatrix}
-\frac{K_0}{K_1}\left\{\left(\frac{V_g(\mathbf{u})}{V_{f0}}\right)^{-K_1}-1\right\}\frac{\partial^2 V_g(\mathbf{u})}{\partial\mathbf{u}^2} + \frac{K_0}{V_{f0}} \left(\frac{V_g(\mathbf{u})}{V_{f0}}\right)^{-K_1-1} \frac{\partial V_g(\mathbf{u})}{\partial\mathbf{u}}\otimes \frac{\partial V_g(\mathbf{u})}{\partial\mathbf{u}}
\end{bmatrix}_{2N\times 2N}.
\end{equation}

\subsection{Extension of the trapped fluid zone on the active contact zone}

\begin{figure}[h]
\begin{center}		
	\includegraphics[width=0.4\textwidth]{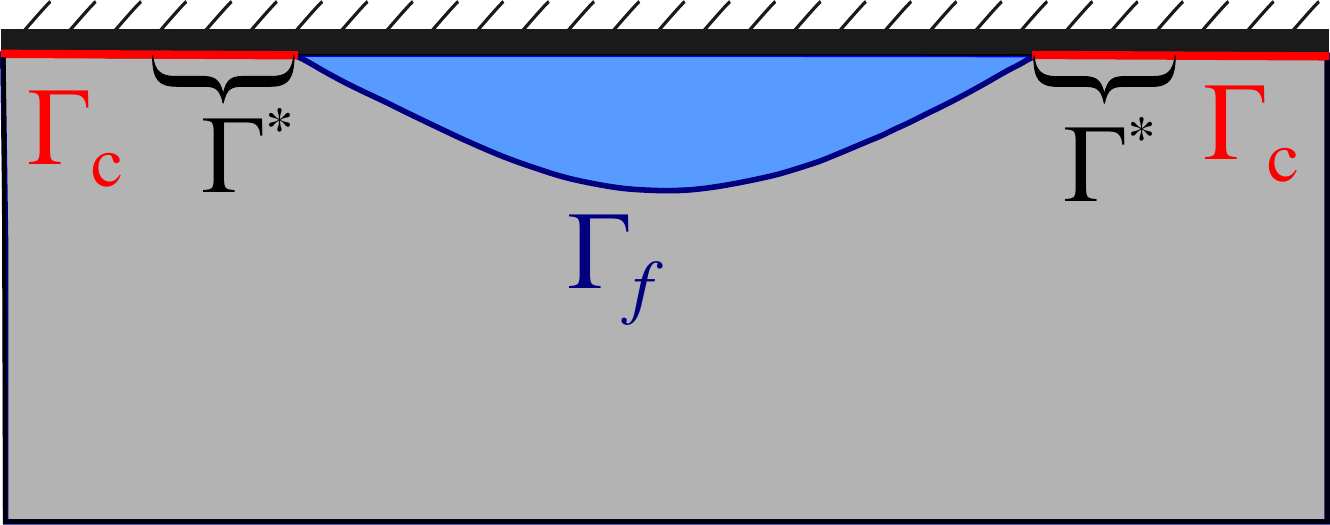} 
	\caption{Extension of the trapped fluid zone $\Gamma_f$ on the active contact zone $\Gamma_c$.}
	\label{fig::fluid_cont_surf}
\end{center}
\end{figure}

Initially we supposed that active contact zone $\Gamma_c$ and trapped fluid zone $\Gamma_f$ are complementary subsets of $\Gamma$ - the whole surface of the deformable body (in the interface): $\Gamma_f \,\cap\, \Gamma_c$ is a set of measure zero, and $\Gamma_f \,\cup\, \Gamma_c = \Gamma$. In accordance with the numerical procedures for solving the coupled problem proposed in the previous subsections, the trapped fluid zone $\Gamma_f$, and, consequently, the number of DOFs of the trapped fluid element must be updated on every iteration of the Newton-Raphson method, which increases the computation time. 
Below we show that in order to simplify numerical calculations we may omit this zone splitting without loss of generality and 
accuracy\footnote{We demonstrate it for the case of the Lagrange multipliers method, however, it may be also generalized to the penalty method.}. 
We make an extension of trapped fluid zone on the active contact zone, i.e. consider fluid pressure on surface $\Gamma_f \cup \Gamma^*$,  where $\Gamma^* \subseteq \Gamma_c$,
see Fig.~\ref{fig::fluid_cont_surf}. The only change we have to take into account is that on $\Gamma^*$ the contact normal pressure will not be equal to the Lagrange multiplier $\lambda_c$ corresponding to contact, but to the difference between the latter and the value of the trapped fluid Lagrange multiplier: $\sigma_n = \lambda_c - \lambda_f \;\text{on}\; \Gamma^*$. Note that $\lambda_c$, which is equivalent to the normal traction, is negative, while $\lambda_f$ represent fluid pressure, which is positive by definition.

In order to prove validity of this extension, we will consider a transformation of the Lagrangian for the coupled system ~(\ref{eq::Lagrangian_coupled}). We will start with substituting the formula for the gap volume~(\ref{eq::gap_volume_integral}) into~(\ref{eq::Lagrangian_coupled}) and obtain:
\begin{equation}
\label{eq::Lagrangian_coupled_rewritten}
\mathcal{L}(\mathbf{u}, \lambda_c, \lambda_f) = \Pi(\mathbf{u}) + \int\limits_{\Gamma_c} \lambda_c \, g(\mathbf{u}) \: d\Gamma_c - \lambda_f \left(\int\limits_{\widetilde{\Gamma}_f} g(\mathbf{u}) \: d\widetilde{\Gamma}_f - V_f \right).
\end{equation}
Let us break the integral over the active contact zone $\Gamma_c$ into two integrals over surfaces $\Gamma^*$ and $\Gamma_c \setminus \Gamma^*$ and consider the following representation of contact Lagrange multiplier $\lambda_c$ on the surface $\Gamma^*$:  $\lambda_c = \lambda_c^* - \lambda_f$, where $\lambda_c^* \leq 0$. Note that this representation is valid only if $|\lambda_c| \geq |\lambda_f|$, which is the case for the problem under study (except for the elasto-plastic case): due to the considered regular wavy profile of the surface and gradual monotonic increase of the external pressure, the  contact pressure must be higher in the contact patches (i.e. in the active contact zone), than in the trapped fluid zone, because otherwise the contact would not be active. 

Therefore, we may write
\begin{align}
\label{eq::Lagrangian_coupled_rewritten_2}
\mathcal{L}(\mathbf{u}, \lambda_c, \lambda_f) =& \Pi(\mathbf{u}) + \int\limits_{\Gamma_c \setminus \Gamma^*} \lambda_c \, g(\mathbf{u}) \: d\Gamma_c + \int\limits_{\Gamma^*} (\lambda_c^* - \lambda_f) \, g(\mathbf{u}) \: d\Gamma_c - \lambda_f \left(\int\limits_{\widetilde{\Gamma}_f} g(\mathbf{u}) \: d\widetilde{\Gamma}_f - V_f \right) \nonumber \\
=& \Pi(\mathbf{u}) + \int\limits_{\Gamma_c \setminus \Gamma^*} \lambda_c \, g(\mathbf{u}) \: d\Gamma_c + \int\limits_{\Gamma^*} \lambda_c^* \, g(\mathbf{u}) \: d\Gamma_c - \lambda_f \left(\int\limits_{\widetilde{\Gamma}_f \cup \Gamma^*} g(\mathbf{u}) \: d\widetilde{\Gamma}_f - V_f \right).
\end{align}
The last formula in~(\ref{eq::Lagrangian_coupled_rewritten_2}) shows that for the coupled problem:
\begin{itemize}
\item the trapped fluid zone $\Gamma_f$ can be extended on a part or on the whole active contact zone $\Gamma_c$ without loss of generality: $\Gamma^* \subseteq \Gamma_c$;
\item if the trapped fluid is in the active state, the value of $\lambda_c^* - \lambda_f$ is equivalent to the normal stress component on $\Gamma^*$.
\end{itemize}


\end{document}